\documentclass[12pt,preprint]{aastex}

\begin{document}
 
\newcommand{\kms}{km s$^{-1}\;$}
\newcommand{\msun}{M_{\odot}}
\newcommand{\rsun}{R_{\odot}}
 
\title{Analysis of Detached Eclipsing Binaries near the Turnoff of the Open
  Cluster NGC 7142}

\author{Eric L. Sandquist\altaffilmark{1}, Matthew Shetrone\altaffilmark{2},
  Andrew W. Serio\altaffilmark{1,3}, Jerome Orosz\altaffilmark{1}}

\altaffiltext{1}{San Diego State University, Department of Astronomy, San
  Diego, CA 92182; {\tt esandquist@mail.sdsu.edu}, {\tt aserio@gemini.edu},
  {\tt jorosz@mail.sdsu.edu}}

\altaffiltext{2}{University of Texas, McDonald Observatory, HC75 Box 1337-L 
    Fort Davis, TX, 79734; {\tt shetrone@astro.as.utexas.edu}}

\altaffiltext{3}{Current address: Gemini Observatory, Southern Operations 
Center, AURA, Casila 603, La Serena, Chile}

\begin{abstract}
We analyze extensive $BVR_CI_C$ photometry and radial velocity measurements
for three double-lined deeply-eclipsing binary stars in the field of the old
open cluster NGC 7142. The short period ($P = 1.9096825$ d) detached binary
V375 Cep is a high probability cluster member, and has a total eclipse of the
secondary star. The characteristics of the primary star ($M=1.288\pm0.017
\msun$) at the cluster turnoff indicate an age of 3.6 Gyr (with a random
uncertainty of 0.25 Gyr), consistent with earlier analysis of the
color-magnitude diagram. The secondary star ($M = 0.871\pm0.008 \msun$) is not
expected to have evolved significantly, but its radius is more than 10\%
larger than predicted by models. Because this binary system has a known age,
it is useful for testing the idea that radius inflation can occur in short
period binaries for stars with significant convective envelopes due to the
inhibition of energy transport by magnetic fields. The brighter star in the
binary also produces a precision estimate of the distance modulus, independent
of reddening estimates: $(m-M)_V=12.86\pm0.07$.

The other two eclipsing binary systems are not cluster members, although one
of the systems (V2) could only be conclusively ruled out as a present or
former member once the stellar characteristics were determined.  That binary
is within $0\fdg5$ of edge-on, is in a fairly long-period eccentric binary,
and contains two almost indistiguishable stars. The other binary (V1) has a
small but nonzero eccentricity ($e=0.038$) in spite of having an orbital
period under 5 d.
\end{abstract}


\section{Introduction}

Of all the methods of determining the ages of
stars (other than the Sun), the method that requires the least theoretical
intervention involves the measurement of the mass and radius of evolved main
sequence stars in detached eclipsing binaries (DEBs). For a group of stars
born at the same time, the most massive (and therefore, the hottest and most
luminous) stars consume their hydrogen fuel the quickest and begin to change
rapidly in size, temperature, and luminosity. The brightest and hottest main
sequence stars remaining thereby indicate the age of the
group.  Unfortunately, observational and theoretical limitations preclude the
measurement of really accurate ages from brightness and color alone ---
uncertainties in distance and interstellar reddening, in the modeling of
convection, and in the conversion from color to surface temperatures are the
most notorious problems.
% Pinsonneault paper has good intro
Masses and radii found from DEBs are unaffected by these uncertainties because
they can be determined using straightforward physics and measured with high
precision.  Mass is a quantity that is {\it explicitly} used in stellar models
that sensitively influences a star's life; radii reveal the evolutionary
state of the stars.

Separately, evolved field DEBs such as AI Phe and TZ For (with well-determined
$M$, $R$, $T_{\rm eff}$, and [Fe/H]; \citealt{andrev}) and photometry of star
clusters (with well-determined distance and [Fe/H]) have been used to
constrain stellar models \citep[e.g.][]{vand}. Ideally though, the
most restrictive constraints will come from DEBs {\it in} star clusters. In
that case, a well-measured DEB can pinpoint the masses of stars at critical
spots in a cluster's color-magnitude diagram, while the rest of the single
cluster members can be used to collectively probe the physics governing the
stars. If we are lucky enough to find {\it multiple} DEBs in a
cluster, the observations would more tightly constrain the wiggle room
available to the theoretical models. A critical aspect of this is to find DEBs
in clusters that have evolved off of the main sequence (in other words,
changed significantly in radius from their main sequence values) because they
break degeneracies involving uncertainties in distance, reddening,
color-$T_{\rm eff}$ transformations, and chemical composition
\citep[e.g.][]{south}.

However, only a handful of DEBs with evolved stars in clusters have been
identified, much less studied in detail. Our previous work on NGC 7142
\citep{sand7142} presented variable star discoveries identified in the process
of characterizing a previously known \citep{ct} eclipsing binary (V375 Cep) at
the cluster turnoff. This paper presents the analysis of the most promising
eclipsing binaries from that study.

% very short summary of results from detection paper... esp metallicity

\section{Observational Material}\label{obs}

The photometry of the binary stars was presented in \citet{sand7142}.
Briefly, the images were obtained at the Mount Laguna Observatory 1m telescope
using a $2048\times2048$ pixel CCD with a field of view about $13\farcm5$ on a
side.  The photometry was originally undertaken for the purpose of
characterizing V375 Cep, but after a second eclipsing binary was identified at
the turnoff, photometric observations were used to determine the ephemeris and
observe eclipses. Since the \citeauthor{sand7142} paper, we obtained
additional observations of the eclipses of V2. These new observations are
listed in Table \ref{datestab}.

We derived light curves from differential photometry using our updated version
of the image subtraction package ISIS \citep{isis}.  One improvement that was
implemented since the \citet{sand7142} paper was that we improved the spline
interpolation routines (from bicubic to Akima splines) that are used to
calculate the point spread function (PSF). The PSF is determined from a subset
of the stars on the frame, and the interpolated PSF is used to weight the
pixels used in the differential photometry. This change resulted in reduced
scatter in the photometry for images with large spatial offsets from the
reference field, or for stars with weaker signal (due to clouds, for example).
The outcomes of this process were time series of magnitudes in the $B$, $V$,
$R_C$ and $I_C$ filters. Stars in the observed field were
calibrated in $BVI_C$ to the standard system using stars from \citet[][retrieved August 2009]{stet}.

Our spectra were obtained at the Hobby-Eberly Telescope (HET) with the High
Resolution Spectrograph (HRS, \citealt{tull}) as part of normal queue
scheduled observing \citep{shetet}.  The configuration of the HRS was chosen
based upon the spectral line widths and strength of the secondary in the first
spectrum taken of each object.  V375 Cep was observed with the configuration
HRS\_15k\_central\_600g5822\_2as\_2sky\_IS0\_GC0\_2x5 to achieve R=15,000,
while V1 and V2 were observed with the
HRS\_30k\_central\_600g5822\_2as\_2sky\_IS0\_GC0\_2x3 to achieve R=30,000.
Both configurations cover 4825 \AA\ to 6750 \AA\ with a small break at
5800 \AA\ between the red and blue CCDs.  Typical exposure times were 900,
1200, and 1680 seconds to achieve signal-to-noise around 50, 45 and 75 at
5800 \AA\ for V375 Cep, V2, and V1, respectively.  The data were reduced using
the echelle package within IRAF\footnote{IRAF is distributed by the National
  Optical Astronomy Observatory, which is operated by the Association of
  Universities for Research in Astronomy, Inc., under cooperative agreement
  with the National Science Foundation.}  for fairly standard bias and
scattered light removal, 1D spectrum extraction, and wavelength calibration.
%We have written a IRAF script that takes the
%1d echelle format output and generates a single long cleaned spectrum ideal
%for cross-correlation.

\section{Analysis}

\subsection{Rotational and Radial Velocities}

\begin{figure}
\includegraphics[scale=0.7]{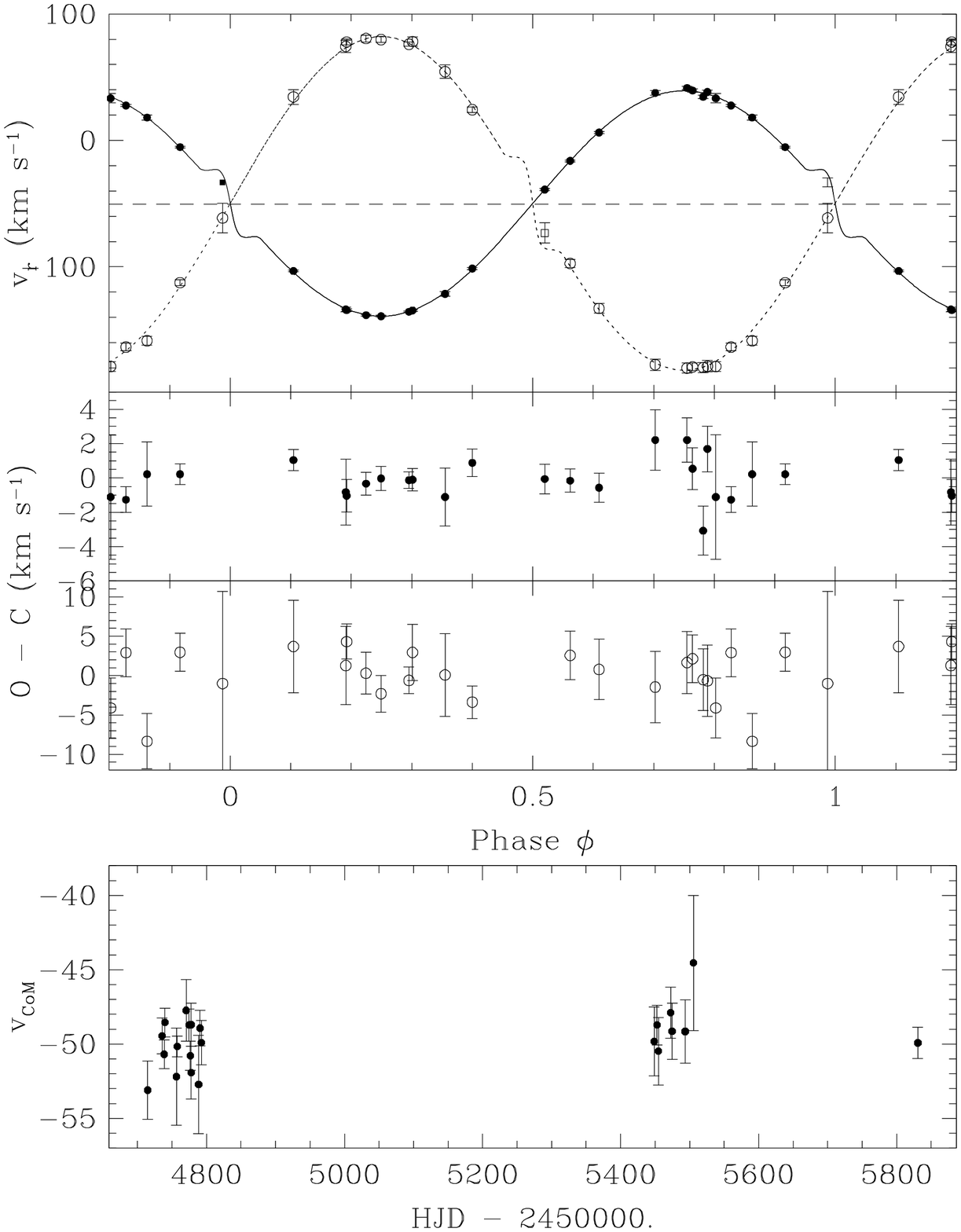}
\caption{{\it Upper panel:} Phased radial velocities for V375 Cep. Model fits
  are shown with solid lines, and the cluster mean radial velocity is shown as
  the flat dashed line. Two observations that were affected by the Rossiter
  effect are shown, but were not used in the fits. {\it Middle panels:}
  Observed minus calculated velocities for the two stars. {\it Lower panel:}
  Calculated center-of-mass radial velocities for the V375 Cep binary as a
  function of time.\label{v375rv}}
\end{figure}

\begin{figure}
\plotone{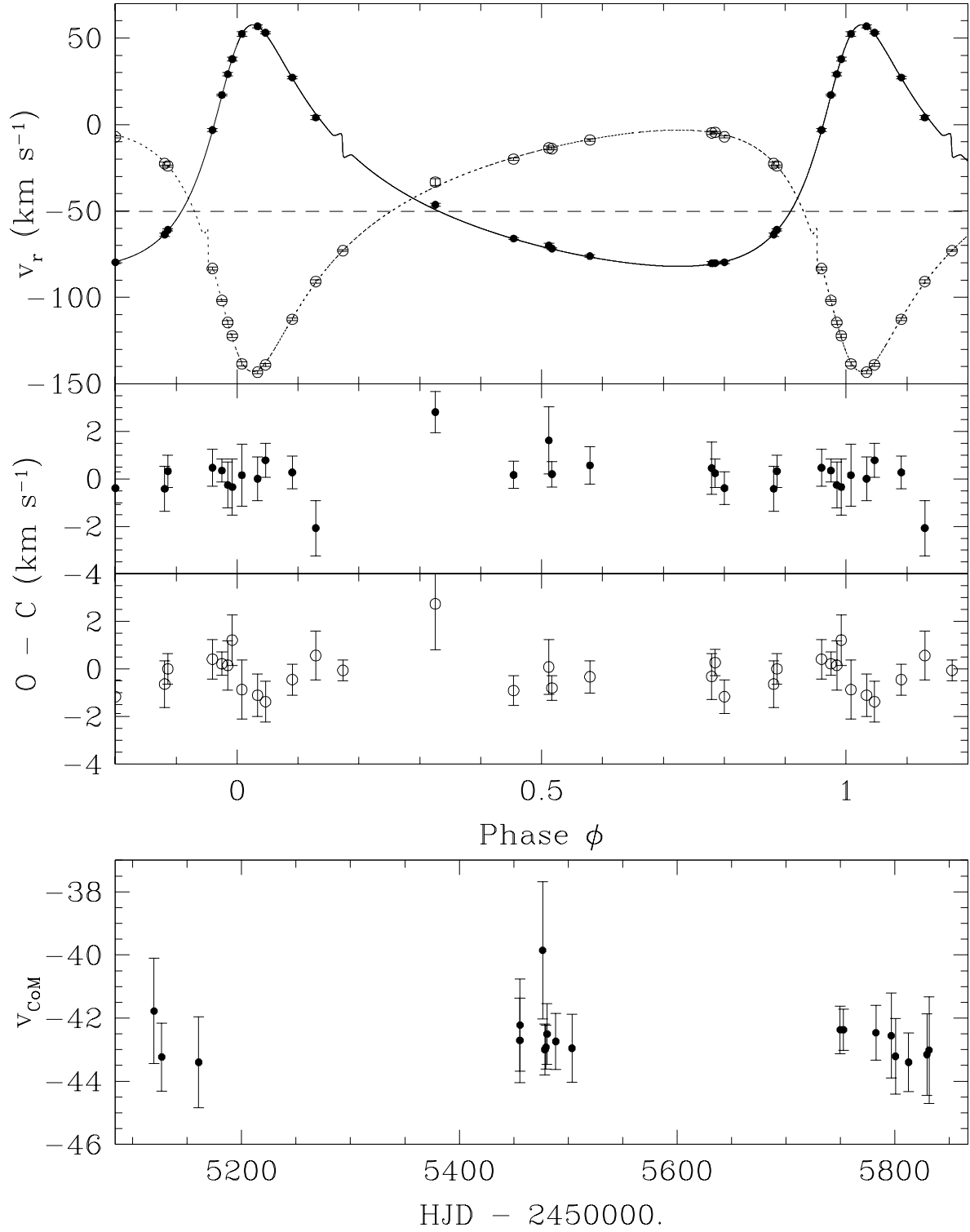}
\caption{Same as in Fig. \ref{v375rv} except for V2.\label{v2rv}}
\end{figure}

\begin{figure}
\plotone{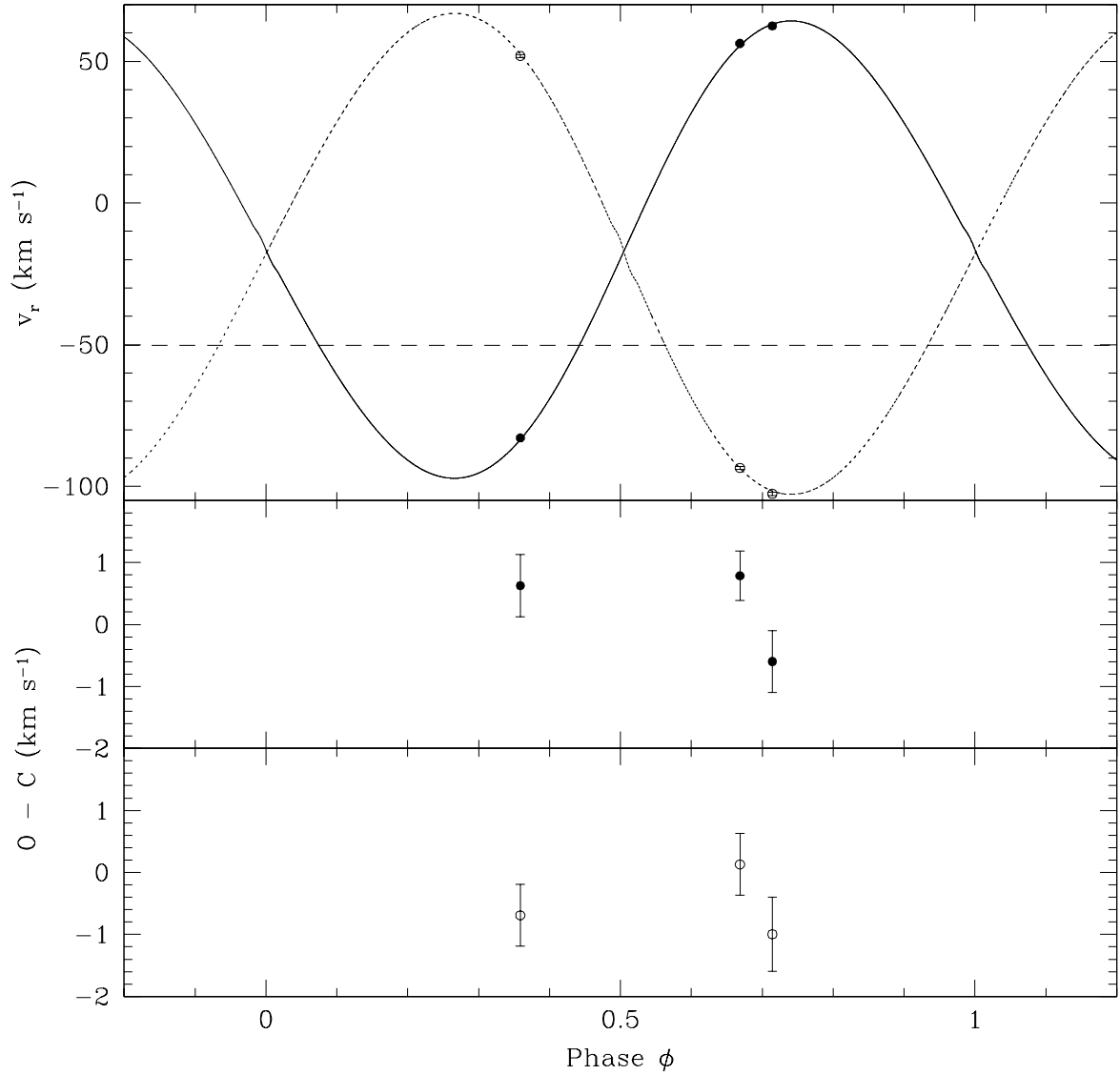}
\caption{{\it Upper panel:} Phased radial velocities for V1. Model fits are
  shown with solid lines, and the cluster mean radial velocity is shown as the
  flat dashed line. {\it Lower panels:} Observed minus calculated velocities
  for the two stars. \label{v1rv}}
\end{figure}

Radial velocities were determined from cross correlation using the IRAF task
{\tt fxcor} with a solar spectrum \citep{hinkle} over the region 4880 to 5750
\AA.  On nearly every night that a cluster star was observed, we also observed
a radial velocity standard, and the difference between the measurement of the
standard and the literature value determined a zero point that was applied to
the final star velocity. The radial velocity corrections only vary slightly (a
few tenths of a km s$^{-1}$) from night to night, but change more
significantly with the instrument configuration and season. If a standard was
not observed on the night of a cluster observation, the correction was taken
by averaging ones from nearby nights. The radial velocities for
the three binaries under consideration here are given in Table \ref{spectab},
while the phased velocity curves are shown in Figs. \ref{v375rv}, \ref{v2rv},
and \ref{v1rv}.

Rotational velocities were determined from clean spectra for both components.
This involved creating a first attempt at a clean spectrum by shifting all of
the spectra to the rest frame for one of the components and then combining the
spectra using a median with a fairly aggresive sigma clip to remove the
spectral features of the other component.  With preliminary A and B component
spectra, we could divide them back into the original spectra (shifted to the
correct velocity) and then repeat the process on these residual spectra files
to generate a second set of cleaned A and B component spectra.  The cleaned
component spectra were then cross-correlated against the solar spectrum, and
the width of the cross-correlation peak measured.  We then generated synthetic
spectra with different rotational velocities and cross-correlated them against
the same solar template spectrum. Finally, we estimated the rotational velocity
by interpolating in the grid of results for cross-correlation peak widths.

\subsection{Abundance Analysis}\label{abund}

Abundance analysis for binaries is a fairly specialized and complicated
endeavor.  To accomplish this, we computed synthetic spectra using the 2010
version of {\sc MOOG} (Sneden 1973) with a line list based largely on the
Kurucz line list\footnote{\tt http://kurucz.harvard.edu/LINELISTS/GFHYPERALL}
but with some lines adjusted to fit a solar spectrum \citep{hinkle}.  {\sc
  MOOG} is able to account for continuum light contributed by a companion star
in the binary.  To minimize the number of free parameters, we used surface
gravities and flux ratios from the binary analysis (because the
uncertainties on these parameters are far smaller than could be obtained
with any spectroscopic analysis) while we let the effective
temperature and metallicity of the model atmospheres vary.

We calculated synthetic spectra on a grid of effective temperatures and
metallicities with steps of 250 K and 0.07 dex, respectively.  We then divided
the observed spectrum into the synthetic spectrum, and in several small
wavelength regions calculated the residuals about a fitted constant value,
where the constant was allowed to change from region to region to compensate
for errors in setting the continuum.  The regions we chose were: 4866-4995
\AA\, 4995-5220 \AA\, 5220-5390 \AA\, 5300-5330 \AA\, and 5390-5620 \AA. The
5300-5330 \AA\ region is given extra weight by being used twice because it
contains a mix of strong lines that increase and decrease with changes in
temperature, making it particularly sensitive to $T_{\rm eff}$. We
interpolated the results between grid points to determine the best
parameters for each region.  The results from the regions were then averaged
together to give a final effective temperature and metallicity, along with an
estimate of the random uncertainties.  For V2 we derived T$_A = 6238\pm52$ K
and [Fe/H]$_A = -0.03 \pm 0.06$, while T$_B = 6276\pm63$ K and [Fe/H]$_B =
-0.12 \pm 0.02$. Because these two stars are found to have identical
characteristics within the uncertainties in the later analysis of the binary,
it is unlikely that the inputs to the spectral modeling (log $g$ and/or the
flux ratios) are responsible for the difference in the metallicities
derived. It is more likely that systematic errors dominate the internal
uncertainties.
For V375 Cep we derive $T_A = 6230\pm 50$ and [Fe/H]$_A = +0.09 \pm 0.02$
(where the quoted uncertainties are errors for the mean), while the B
component was too weak to yield useful results. If we consider systematic
uncertainties due to inputs for the spectroscopic analysis (such as oscillator
strengths and microturbulence), the uncertainties are larger. We estimate that
the overall uncertainties are 100 K for $T_A$ and 0.05 dex for [Fe/H]$_A$. The
abundance derived by \citet{spec} for NGC 7142 giants is $+0.14\pm0.01$.  Our
metallicity for V375 Cep is thus within $2\sigma$ of the \citeauthor{spec}
value while our metallicity for the V2 system suggests that it may be a
non-member.

\subsection{Cluster Membership}\label{mem}

Membership determinations for a poorly-studied cluster like NGC 7142 can be
fairly difficult. To date, proper motions have only been published for a few
stars in the cluster field \citep[e.g.][]{baumg}. In addition, there have only
been a relatively small number of high precision radial velocity
measurements. \citet{jaco1} identified 6 cluster stars out of a sample of 17
and found an average radial velocity of $-48.6\pm1.1$ km s$^{-1}$, while \citet{spec}
found $-50.3 \pm 0.3$ \kms from higher resolution spectra of 4 of the same
candidate members. \citet{sand7142} observed three red clump star candidates,
and found that one had a velocity consistent with these averages, but two had
velocities of $-43.9$ and $-44.0$ km s$^{-1}$. Looking more
carefully at the photometry for these stars, the two stars with the higher
velocities are fainter than other candidates in the 2MASS $K_s$ band by more
than 0.2 mag, but bluer in the $(J-K_S)$ color. This could indicate that these
are foreground giant stars.

From the binary modeling discussed later, we find a system velocity $\gamma =
-17.2$ km s$^{-1}$ for V1, which unambiguously rules out cluster
membership. By comparison, the system velocity for V375 Cep was found to be
$\gamma = -49.86 \pm 0.05$ km s$^{-1}$, in very good agreement with the mean
values found by the two high-resolution spectroscopic studies of the
cluster. We therefore judge V375 Cep to be a very likely cluster member.

The fit for V2 returns a system velocity ($\gamma=-42.57\pm0.02$ km s$^{-1}$)
that falls near the mean cluster value, but about 7-8 \kms
higher. Although there has not been extensive enough proper motion or
radial velocity survey of cluster stars to determine a reliable velocity
dispersion for NGC 7142, the dispersion is expected to be $\lesssim$
1 \kms for bound clusters with typical masses and radii \citep{ocsigma}, and
most old open clusters do seem to have radial velocity dispersions of that
size (NGC 188, \citealt{geller188}; NGC 6819, \citealt{hole}; Berkeley 32,
\citealt{randichbe32}). Thus, V2 is {\it unlikely} to simply be in the wing of
the cluster radial velocity distribution.
The effects of a tertiary on a long period orbit could potentially produce
this difference between the presently measured system velocity and the cluster
mean, and we examine that possibility in more detail in the next subsection. A
strong gravitational interaction within the cluster could also give a cluster
member enough energy to escape.

We can examine other information, such as projected sky position and CMD
position, that provides circumstantial evidence. \citet{janes} found an
``effective'' radius of $4\arcmin$ for the cluster, which roughly corresponds
to 1.5 times the $\sigma$-width of a Gaussian fitted to the
stellar distribution. V375 Cep is projected $2\farcm9$ from the cluster
center, nonmember V1 is $3\farcm9$ from center, and V2 is approximately
$4\farcm8$ from center. Once again V2 has a lower likelihood of cluster
membership, but this could also be related to its large velocity relative to
other cluster members.

As can be seen in Fig. \ref{decompose}, the system photometry and decomposed
optical photometry of V375 places it firmly within the main sequence band. The
colors of V1 are very similar to those of the other EBs despite indications
that both component masses are lower than the primary masses of the other
binaries (see \S \ref{mr}), which is consistent with the smaller reddening of a
foreground object.

The system photometry for V2 is brighter and slightly bluer than cluster
turnoff stars, but when the photometry is decomposed, only the color is
slightly discrepant --- the stars are at the blue edge of the distribution at
the turnoff in the CMD, which could be explained by lower-than-average
reddening.  They are slightly brighter than the primary star in V375 Cep in
optical bands. In the 2MASS bandpasses, they are approximately the same
brightness, assuming that the individual stars are about 0.75 mag fainter than
the combined photometry of V2 and the primary of V375 Cep is about 0.15 mag
fainter than the combined photometry of that binary (roughly consistent with
the $I$-band secondary eclipse depth).  If V2 is a member of the cluster
having lower reddening than V375 Cep, it is consistent that the primary of
V375 Cep is brighter relative to the two stars of V2 in the 2MASS bands.

To summarize, we judge V1 to be nonmember based on its radial velocity.
For V375, the argument for cluster membership is much stronger than it is for
V2. None of the information we have available unambiguously supports V2
membership. The age determination for V2 more definitively argues against
cluster membership, however, and that will be discussed in \S \ref{v2mr}.

\subsection{Search for Tertiary Stars}

Because there are now numerous examples of known triple systems in open
clusters \citep[a partial list includes][]{mermay,mermil,alencar,s1082,liu,jeffr}, it is worth checking whether the influence of tertiary stars can be
detected. If a tertiary star is massive or bright enough, it can affect the
models of an eclipsing binary enough to produce significant systematic errors
in the measured characteristics of the eclipsing stars.  None of the binaries
we discuss here had a third set of detectable lines in our spectra, but with a
long enough baseline of observations, photometric methods (such as eclipse
timing) or spectroscopic methods (such as center-of-mass motion) can reveal
tertiaries via their effects on the eclipsing binary.  Table \ref{mintab}
gives our measurements of the times of eclipse minimum for the binaries. Due
to the relatively small number of radial velocity and eclipse minimum
observations for V1, the detection of a tertiary star's effects is unlikely,
so we do not discuss it here.

The radial velocity results for V2 are shown in Fig. \ref{v2rv}, assuming the
best fit mass ratio $q=1.001$. This binary shows more than a 7 \kms offset
from the cluster mean velocity ($-50.3 \pm 0.3$ \kms from 4 stars;
\citealt{spec}), which could potentially result from the action of a tertiary
star. However, the center-of-mass velocities did not vary significantly over
three seasons and more than a 700 d interval of observations, and there
is no sign of variations in eclipse timing.

\begin{figure}
\plotone{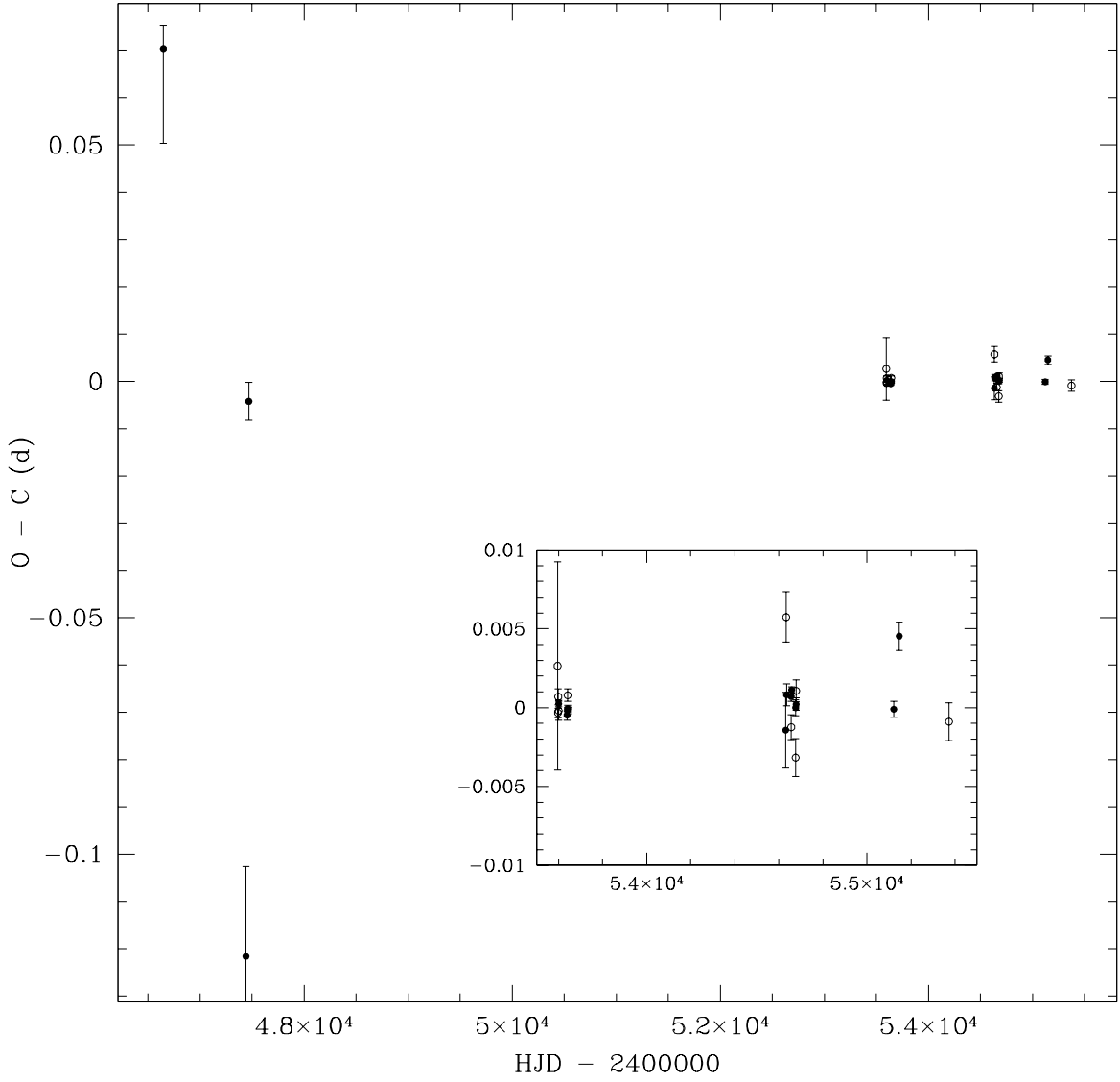}
\caption{Observed time of eclipse versus prediction of the linear ephemeris
  for V375 Cep from our photometric observations. Primary eclipses are shown
with $\bullet$, and secondary eclipses are shown with $\bigcirc$.\label{omc}}
\end{figure}

The lower panel of Fig. \ref{v375rv} shows the measured center-of-mass
velocities for V375 Cep, assuming the best fit mass ratio $q=0.676$ from the
binary models. There does not appear to be evidence of significant motion
during the three seasons (covering more than 1100 days) that we observed the
system. Our own eclipse observations for V375 Cep cover a period of almost
1800 d. For the finely sampled light curves from our study, we used the method
of \citet{kwee} to determine times of minima and the errors, and we show a
comparison of those times with a best-fitting linear ephemeris in
Fig. \ref{omc}. We include in Table \ref{mintab} our best estimates of
eclipse minima from the published observations of \citet{ct} and \citet{see}
to improve the accuracy of the ephemeris and test for the possibility of a
nonlinear ephemeris over the 27 y baseline.

The earlier observations were discussed in \citet{sand7142}.  Most of the
observations by \citet{ct} agree well with our phased light curve and they had
observations in and out of eclipse on the night of one eclipse. Using our
model light curves in $BV$, we fit their data in order to derive an
approximate time of minimum. One additional observation in $V$ on a different
date also appears to have fallen near an eclipse minimum. \citeauthor{ct}
shifted the photometric zeropoints of each of their frames to be consistent,
so the relative photometry of the single observation should be approximately
correct. The interval between this and the nearest primary eclipse is about 14
orbital cycles, but implies a period of about 1.9164 d, which is significantly
different than we find in our more recent observations.

\citet{sand7142} concluded that data from \citet{see} was not of
sufficient quality to test for nonlinearities in the
ephemeris. \citeauthor{see} quoted fairly large uncertainties (0.05 mag) on
their measurements, and the shape and depth of the observations on one night
(HJD 2446650) that appeared to contain a primary eclipse egress were
inconsistent with our model light curves. This may be due to their use of
photographic plates as the recording medium.

We conclude that there is a possibility of eclipse timing variations for the
V375 Cep system, but the fact that it has been well-behaved during the time
covered by our own eclipse observations and radial velocity measurements makes
the existence of a tertiary less probable.

\subsection{Reddening, Stellar Photometry, and Temperature Estimates}

In order to use the photometry for temperature estimates for the stars, we
need to have a measurement of the reddening. \citet{sand7142} derived a
reddening value [$E(B-V)=0.32\pm0.06$] via a comparison of the red clump stars
in NGC 7142 with those of M67. We revisit that estimate here by examining how
the difference in median clump magnitudes between the two clusters changed
with filter. M67 and NGC 7142 have similar ages and metallicities, and have
values in ranges where small differences have minimal effects on the
photometry of the red clump \citep{girardi,groc}.

We have, however, calculated theoretical corrections for intrinsic differences
in clump magnitude from \citet{girardi} models in order to make the reddening
determination more precise.  At constant age, the higher metallicity of NGC
7142 ($\Delta$[Fe/H]$=0.14$) is theoretically expected to make the clump
magnitude brighter in 2MASS infrared filters (by 0.05 mag in $K_s$), but
increasingly fainter at bluer wavelengths, reaching almost 0.13 mag in $B$.
At constant metallicity, the larger age of M67 is expected to make the red
clump fainter in all filters by approximately 0.03 mag
\citep{girardi,groc}.

\begin{figure}
\includegraphics[scale=0.7]{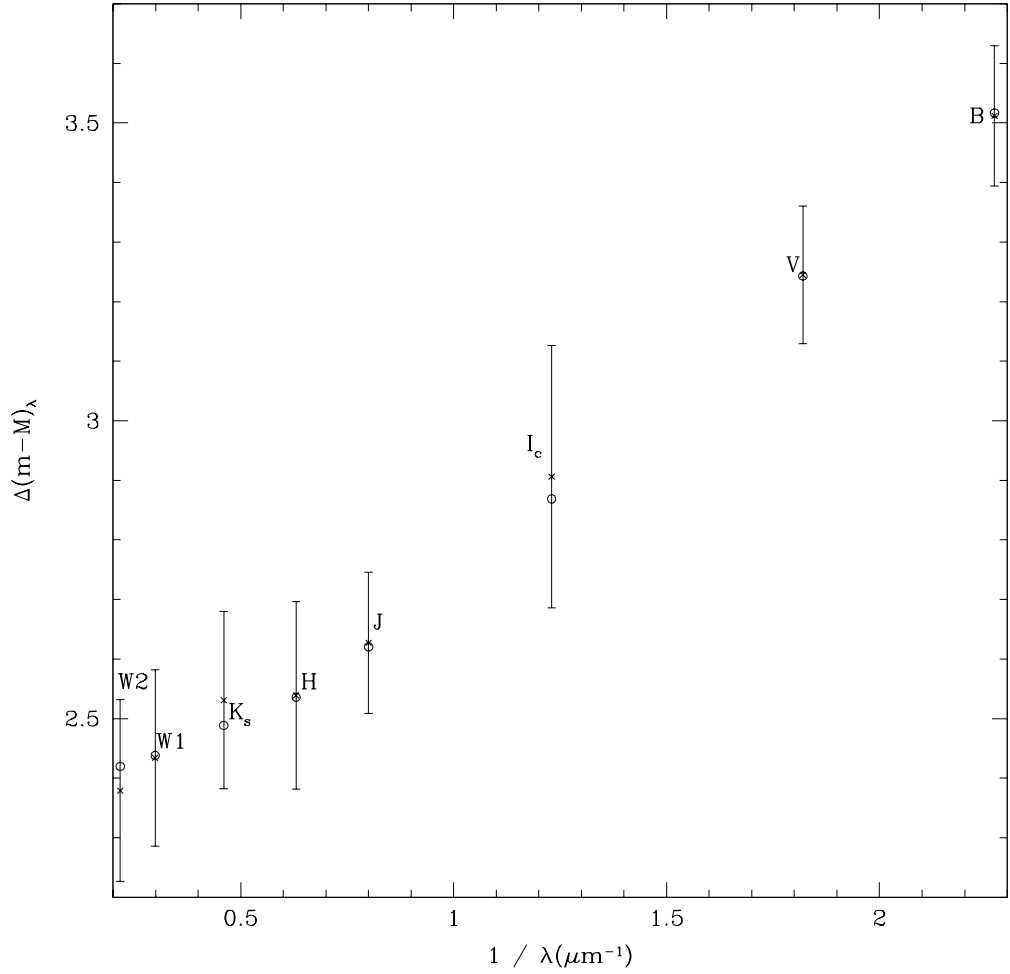}
\caption{Difference in median red clump star magnitudes between the clusters NGC 7142 and M67 as a function of filter. Observed values are shown with $\times$ symbols, while the best fit values are shown with $\circ$ symbols.
$\times$.\label{clumpred}}
\end{figure}

We made use of our own optical photometry along with 2MASS \citep{2mass} and
WISE \citep{WISE} infrared photometry to simultaneously derive the differences
in true distance moduli [$\Delta (m-M)_0 = 2.45^{+0.11}_{-0.07}$] and optical
depths ($\Delta \tau_1=0.278\pm0.053$) between the two clusters, assuming an
extinction law based on the study by \citet{mccall} with \citet{ccm} used
to extend predictions to the WISE filters. $\Delta (m-M)_0$ is primarily
determined by observations in the infrared where the extinction is small,
while $\Delta \tau_1$ is constrained by the variation in the extinction from
filter to filter. The fit is shown in Fig. \ref{clumpred}. The
uncertainties on each measurement are based on uncertainties on the medians of
each clump (dominated by the uncertainties for NGC 7142), and the goodness of
fit was calculated using a $\chi^2$ algorithm. The uncertainties in $\Delta
(m-M)_0$ and $\Delta \tau_1$ were derived from the ranges covered by fits that
were within 1 of the minimum value. Using the well-determined distance modulus
\citep[$(m-M)_0=9.60\pm0.03$;][]{sandquist04} and reddening
\citep[$E(B-V)=0.041\pm0.004$;][]{taylor} for M67, we find $(m-M)_0 =
12.05^{+0.11}_{-0.09}$ and $E(B-V)=0.29\pm0.05$ for NGC 7142. Because there
appears to be a significant amount of differential reddening in the cluster,
infrared colors should be employed when possible.  We discuss below the
characteristics of each system that we are able to exploit to produce
temperature estimates from the photometry of the binary stars.

\begin{figure}
\includegraphics[scale=0.7]{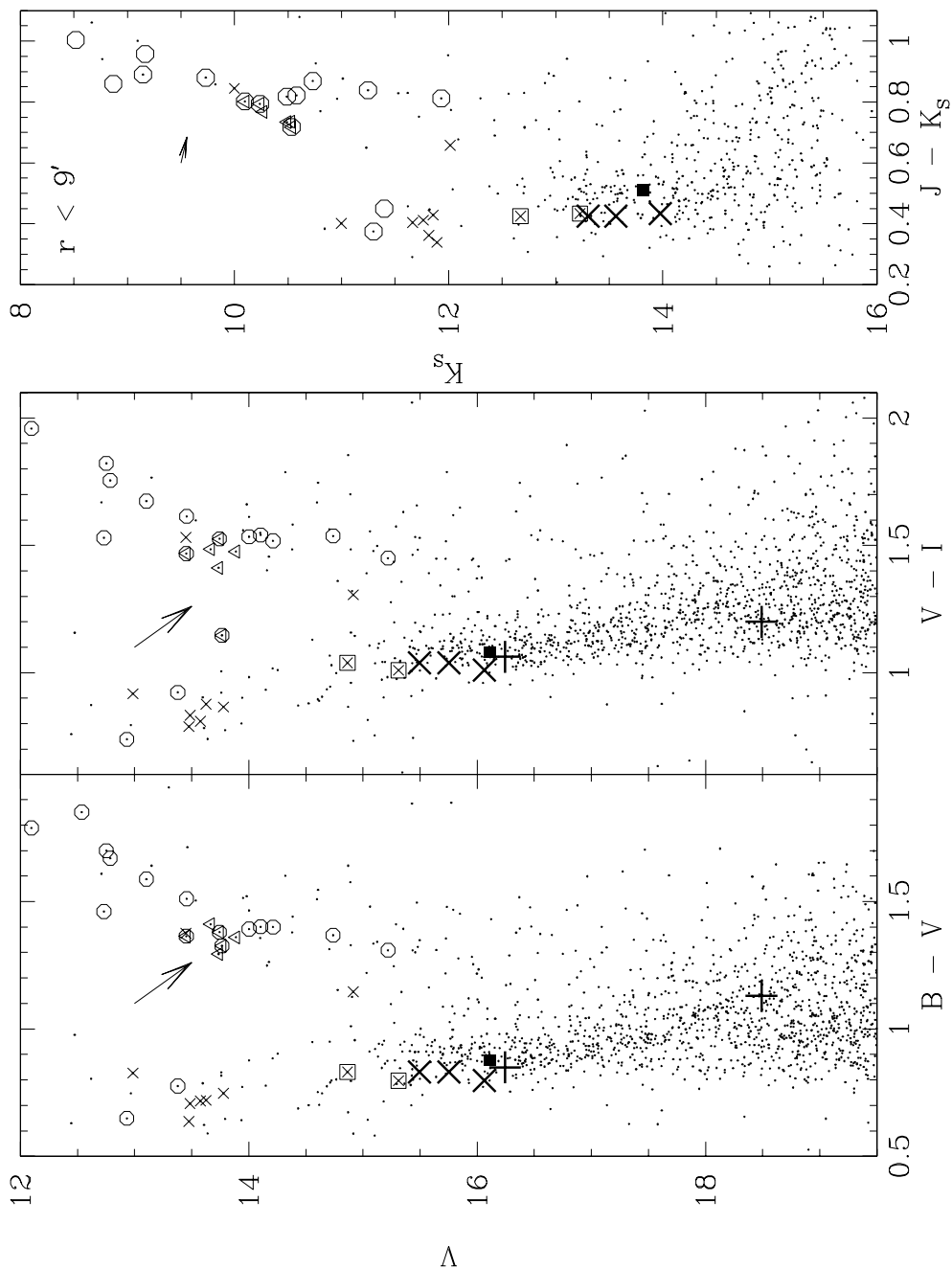}
\caption{Color-magnitude diagrams for NGC 7142 with the system photometry
  ($\blacksquare$ for cluster member V375 Cep, $\sq$ for nonmembers) and
  binary star components identified ($+$ for members, $\times$ for
  nonmembers).  Probable cluster members (identified from spectroscopic radial
  velocities) are shown with small open circles, and nonmembers are shown with
  small $\times$.\label{decompose}}
\end{figure}

The characteristics of the binary systems make it possible to obtain good
estimates of the colors of the component stars.  Table \ref{phottab} lists the
photometry for the binary systems and their components, and
Fig. \ref{decompose} shows their positions in the CMD. 

For V2, the two eclipses are deep and very nearly the same in depth, which
supports our later results (\S \ref{v2comb}) that the two stars are nearly
identical in all of their major characteristics. As a result, the system color
is an excellent representation of the star colors. (The components are
therefore about 0.75 mag fainter than the combined photometry.) Later results
also indicate that the binary is probably not a member of the cluster, and is
slightly behind the cluster. To get a temperature estimate, we therefore use
the infrared colors of the binary and assume that the reddening and
metallicity of the binary are close to that of NGC 7142. Using these
assumptions, we find $T_{\rm eff} = 6150\pm200$ K from the $(J-K_S)$ color
using the color-temperature calibration of \citet{casa}.
The temperatures derived from optical colors are consistent with this
estimate, but are much more uncertain due to reddening uncertainties.

The main difficulty for V1 is that it appears to be a foreground system, and
so the cluster metallicity and reddening do not apply. However, we can derive
fairly accurate temperature estimates if we note that the stellar temperatures
only differ by a little under 2\% according to later models (see \S
\ref{v1comb}), so that the system color will be a fair representation of the
colors of the components. (The luminosities of the stars differ, however, so
we have used the results of the binary models to determine the fraction of the
flux contributed by each star. The estimates of the component magnitudes are
given in Table \ref{phottab}, and the results are plotted in
Fig. \ref{decompose}.) We can minimize the effects of reddening uncertainties
by using infrared colors. A reddening approximately equal to the mean cluster
reddening gives an upper limit to the average temperature of about 6120
K. Given the slightly super-solar masses of the stars (again, see \S
\ref{v1comb}), the Sun's temperature provides us with a lower temperature
limit. Temperature uncertainty due to the unknown metallicity is likely to be
small (a few 10s of K) as long as the stars have near-solar abundances. Based
on these arguments we constrain the {\it primary} (hotter) star temperature to
be between 5850 K and 6250 K.

In the case of V375 Cep, we can see a period of totality in the secondary
eclipse, so that the light from the secondary star can be precisely
disentangled from that of the primary. The errors on the secondary star
photometry in this case are calculated from
\[ \sigma^2(m_2) = \sigma^2(m_{12}) + \frac{\sigma^2(\Delta m_2)}{(10^{(\Delta m_2 / 2.5)} -1)^2}\]
where $m_2$ is the secondary magnitude, $m_{12}$ is the binary magnitude, and
$\Delta m_2$ is the secondary eclipse depth. For shallow eclipse depths, the
factor in the denominator of the second term amplifies the uncertainty
considerably.

We do not have measurements of the secondary eclipse depths in 
infrared filters for V375 Cep, so we resort to optical/near-infrared colors.
These imply $T_A = 6080\pm170$ K for the primary star, again using the
\citet{casa} calibration. For comparison, we obtained a temperature
$6230\pm100$ K, [M/H]$=0.09\pm0.05$, and [$\alpha$/Fe]$=0.0\pm0.15$ from our
spectroscopic analysis in \S \ref{abund}. There is greater uncertainty for the
secondary star resulting from the uncertainties in the photometric
deconvolution, but the most certain determination using the $(V-I)$ color puts
its temperature at about $5050\pm180$ K.

\section{Analysis of the Detached Eclipsing Binaries}

To model the radial velocities and photometry from MLO, we used the Eclipsing
Light Curve code \citep[hereafter ELC; ][]{elc}. ELC is a versatile code, and
we briefly describe the most relevant features here. ELC is capable of fitting
for a number of different binary star parameters depending on the situation,
and the quality of the model fit was judged by an overall $\chi^2$.  The
minimum value can be sought using a genetic or Markov chain Monte Carlo
algorithm. After an initial optimization run, the error bars on the data were
scaled to return a reduced $\chi^2_\nu = 1$ for each type of measurement.  The
reason for this is that the magnitudes of the estimated measurement
uncertainties affect the uncertainties in the derived parameters through the
$\chi^2$ values. So to maximize the reliability of the parameter uncertainty
estimates, we use observational uncertainties that are reflective of scatter
around a best fit model. The quoted parameter uncertainties are based on the
range of values that produce a total $\chi^2$ within 1 of the minimum value
\citep{avni}.

For light curve models, we made use of ELC's ability to describe
center-to-limb intensity variations using either analytic limb darkening laws
or model atmospheres. When using analytic limb darkening, we chose a quadratic
law with two coefficients for each star, where the coefficients are expected
to be dependent on surface temperature, gravity, and composition. Because of
the possibility that systematic errors might be introduced through the use of
incorrect limb darkening coefficients, we selected one coefficient ($x$) for
each star from ATLAS atmospheres \citep{claret} and fit for the other
coefficient ($y$). The effects of systematic errors in one coefficient can be
mitigated by such a fit because the coefficients tend to be correlated
\citep{southworth}. Alternately, we used PHOENIX model atmospheres
\citep{hauschildt} to describe the variation of emitted intensity with
emergent angle, which removes the need to assume limb-darkening coefficients.
However, systematic errors could still be introduced to our binary models if
the $T_{\rm eff}$ values we used are incorrect or if there systematics in the
atmosphere models.

\subsection{V375 Cep}

\begin{figure}
\includegraphics[scale=0.8]{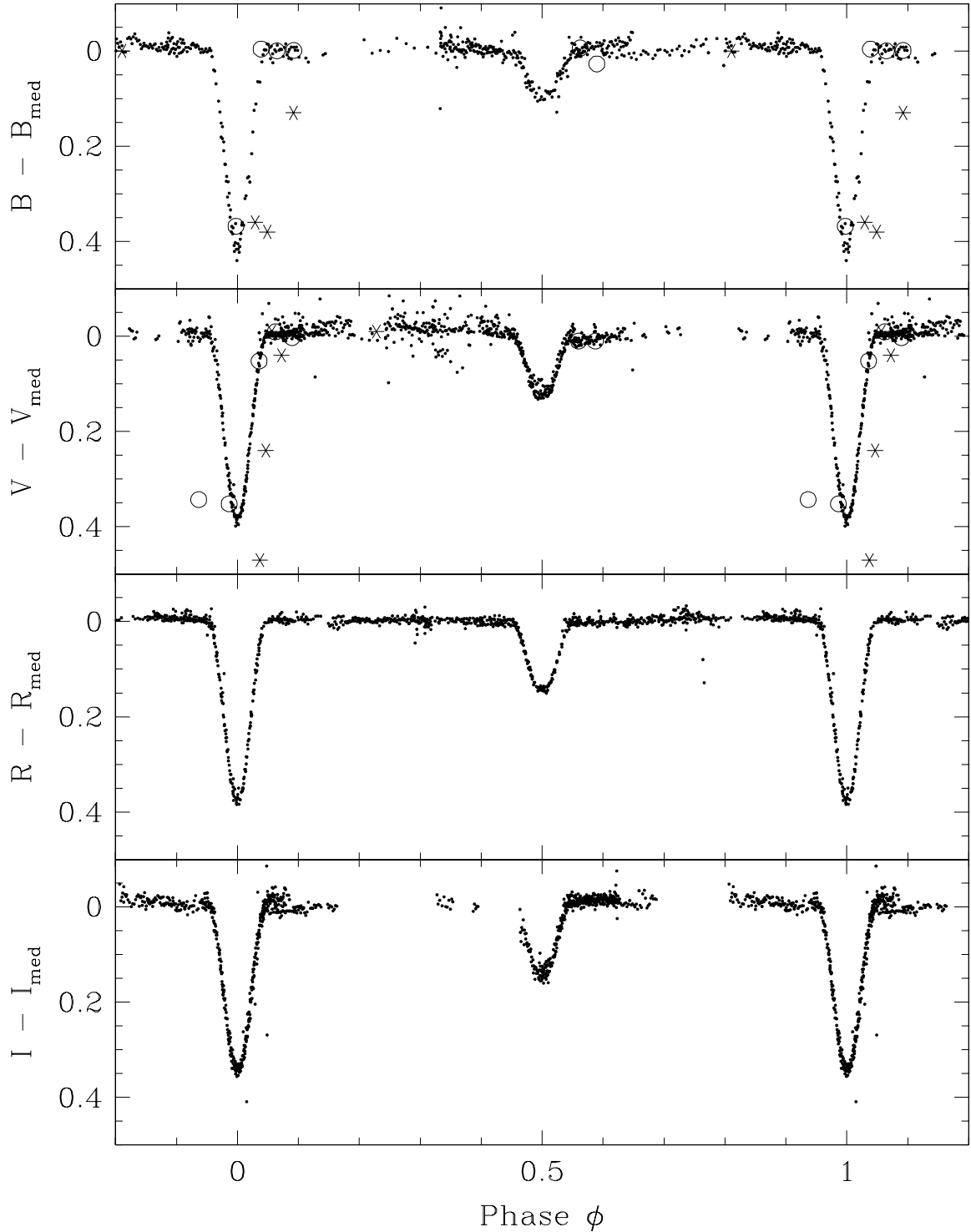}
\caption{$BVRI$ phased light curves for the detached eclipsing binary V375
  Cep. Open circles indicate measurement made by \citet{ct} and asterisks are
  measurements made by \citet{see} phased to our ephemeris.
\label{v375lcs}}
\end{figure}

The light curves (with primary and secondary eclipses of different depths, as
seen in Fig. \ref{v375lcs}) and radial velocities for V375 both implied from
the start that the mass ratio for this system was likely to be significantly
different from 1. This led us to believe that the system's light was dominated
by one of the stars, and that star therefore resided close to the cluster
turnoff. The decomposed photometry shown in Fig. \ref{decompose} confirms
this, and makes the primary star an excellent candidate for constraining the
cluster age if it is a member.

Another notable feature of the light curves for this system is the small
amount of out-of-eclipse light variation. This can be seen especially in the
$R_C$ light curve, which generally had the images with the highest
signal-to-noise ratio and also had the best coverage of out-of-eclipse
phases. In spite of the rather short orbital period, effects due to the
distortion of the stellar surfaces by the other star are barely
discernable. Most of the scatter in other filter bands comes from observations
made during poor weather conditions. However, there is a hint that stellar
activity might be producing some night-to-night variations in $B$. More
observations would be needed to confirm this.

\subsubsection{Radial Velocity Modeling}

Because this binary appears to be circularized and there is little or no
out-of-eclipse light variation, the light curves do not effectively constrain
the mass ratio of the stars. Therefore we modeled the radial velocities
separately from the light curves. In the radial velocity models, we fit for
the velocity semi-amplitude of the primary star $K_A$, the mass ratio
$q=M_B/M_A$, and the system velocity $\gamma$ as parameters. We did an
experiment where we allowed the system velocities to differ for the two
stars (to allow for differences in gravitational redshift or convective
blueshift resulting from their differing evolutionary states), but found
a difference of only 0.1 \kms. This negligibly affected the derived masses.

The initial estimates of the velocity uncertainties for the two stars were
scaled separately to return reduced $\chi^2$ values of 1. After scaling, the
typical uncertainties were $0.5 - 2$ \kms for the primary and $2-4$ \kms for
the secondary. Once the light curves were modeled, the orbital inclination $i$
was used in a final modeling run to derive the stellar masses. The results are
given in Table \ref{chartab}.

\subsubsection{Light Curve Modeling}

In our fits of the light curves, we separately ran models using a quadratic
limb darkening law and using PHOENIX model atmospheres. When using model
atmospheres, the limb darkening is fully described, and so we fitted for 6
parameters: orbital period $P$, time of primary eclipse $t_0$, inclination
$i$, ratio of the primary radius to average orbital separation $R_A / a$,
ratio of radii $R_A / R_B$, and temperature ratio $T_B / T_A$. In the models
using a limb darkening law, we fit for one coefficient of the limb darkening
law for each star in each filter, thereby adding 8 additional
parameters.  In both cases,
the results of the radial velocity fits (specifically, $K_A$ and $q$) and the
spectroscopic temperature of the primary star $T_A$ were input as constrained
values along with their uncertainties. This means their values were allowed to
vary, but models incur a $\chi^2$ penalty as the value deviates more and more
from the constraint.

Although the out-of-eclipse light curve variations are small and indicate that
there is little tidal distortion of the stellar surfaces, we find that if we
{\it assume} that the two stars are spherical our radius measurements end up
systematically higher by about 1\%. This appears to be because the small
out-of-eclipse variations are taken to be part of the eclipses in the fits.
When we allow for nonsphericity though, we find good consistency between our
model atmosphere and limb-darkening law fits.

In a short period binary such as V375 Cep, stellar activity can produce
variations in the light levels. For this reason, we opted to shift nights with
eclipse observations to a common zeropoint (as determined by out-of-eclipse
observations on the same night) in order to remove possible spot
modulation. These shifts were never more than 0.025 mag, and were most
frequently less than 0.015 mag. We did not do the same for nights when the
system was observed completely out of eclipse so that we did not remove the
signature of non-spherical stars.
We will come back to
the issue of whether more stellar activity should be present in \S
\ref{v375disc}.

\subsection{V2}

\begin{figure}
\plotone{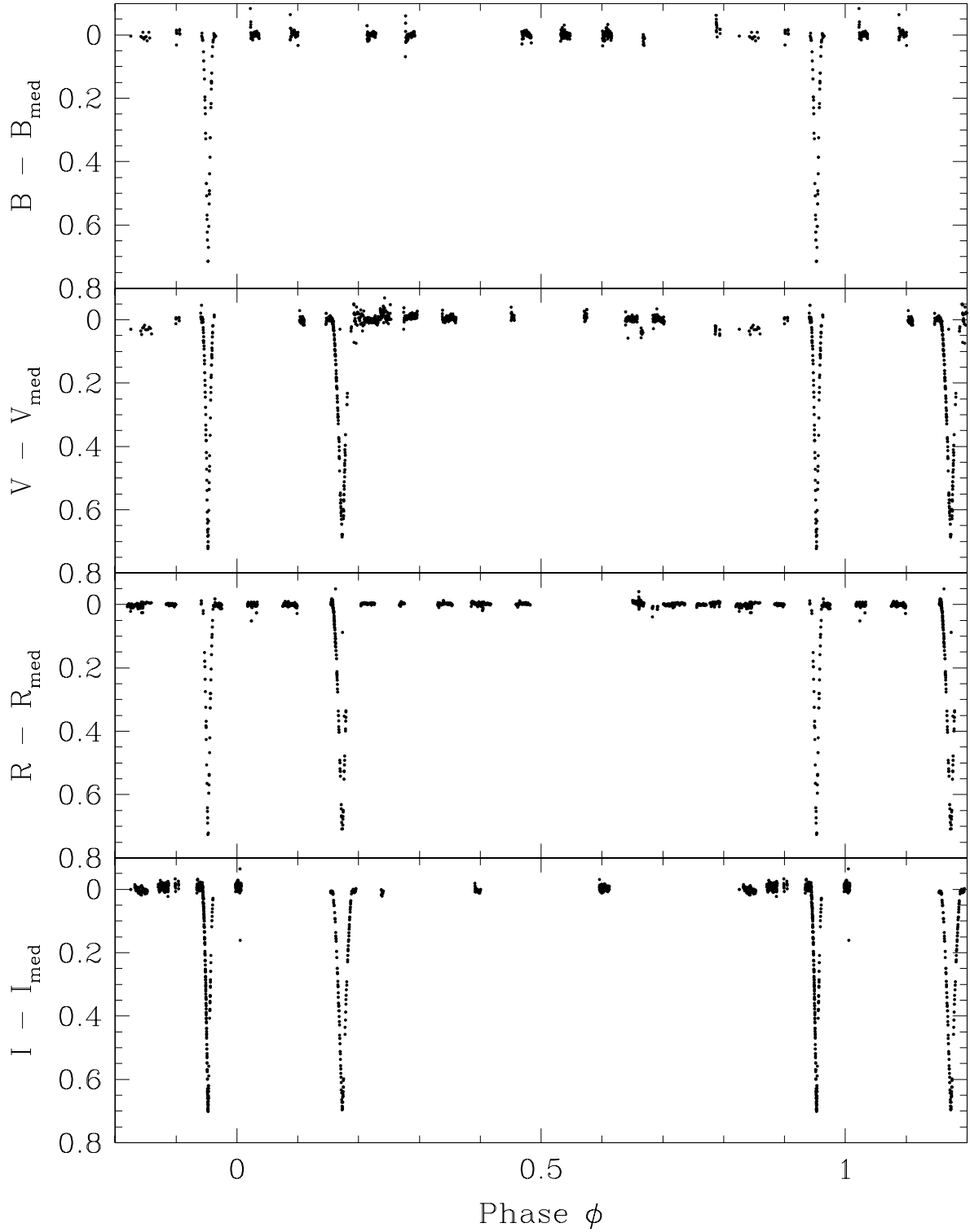}
\caption{$BVRI$ phased light curves for the detached eclipsing binary V2.
Phase $\phi=0$ corresponds to periastron.
\label{v2lcs}}
\end{figure}

The light curves of V2 show two very deep ($0.7-0.8$ mag) eclipses per cycle
(see Fig. \ref{v2lcs}), and two components are very clearly seen in spectra of
the system.  
The separation of the eclipses in phase ($\Delta \phi
= 0.2206$) and the much longer duration of the shallower eclipse conclusively
show that the system has a substantial eccentricity.

\subsubsection{Combined Radial Velocity and Light Curve Modeling}\label{v2comb}

For eccentric binaries, both the radial velocities and the light curves
contain information on the {\it orbits} of the two stars, so it is more
important to model the two datasets simultaneously. 
As we did with V375 Cep, we
scaled the errors for each dataset (photometry by filters, radial velocities
for each component) separately to produce a reduced $\chi^2_\nu$ value near 1.

For a combined run with model atmospheres, we fitted the binary with a set of
12 parameters: orbital period $P$, time of periastron $t_0$, velocity
semi-amplitude of the primary star $K_A$, mass ratio $q$, system velocity
$\gamma$, eccentricity $e$, argument of periastron $\omega$, inclination $i$,
ratio of the stellar radii to average orbital separation $R_A / a$ and $R_B /
a$, primary star temperature $T_A$, and temperature ratio $T_B / T_A$. When
using a quadratic limb darkening law, we forced the fitted limb darkening
coefficients to be the same for both stars due to the indications that the
star temperatures, masses, and radii were nearly identical.  Generally when
limb darkening coefficients are fitted, they are not tied to the stellar
temperatures. In our case, when we allowed the coefficients to vary
independently, the fits converged on values that were signficantly different
for the two stars.  The most likely reason is that systematic trends in the
eclipse light curves were presenting $\chi^2$ incentives for the coefficients
to differ.

\begin{figure}
\includegraphics[scale=0.8]{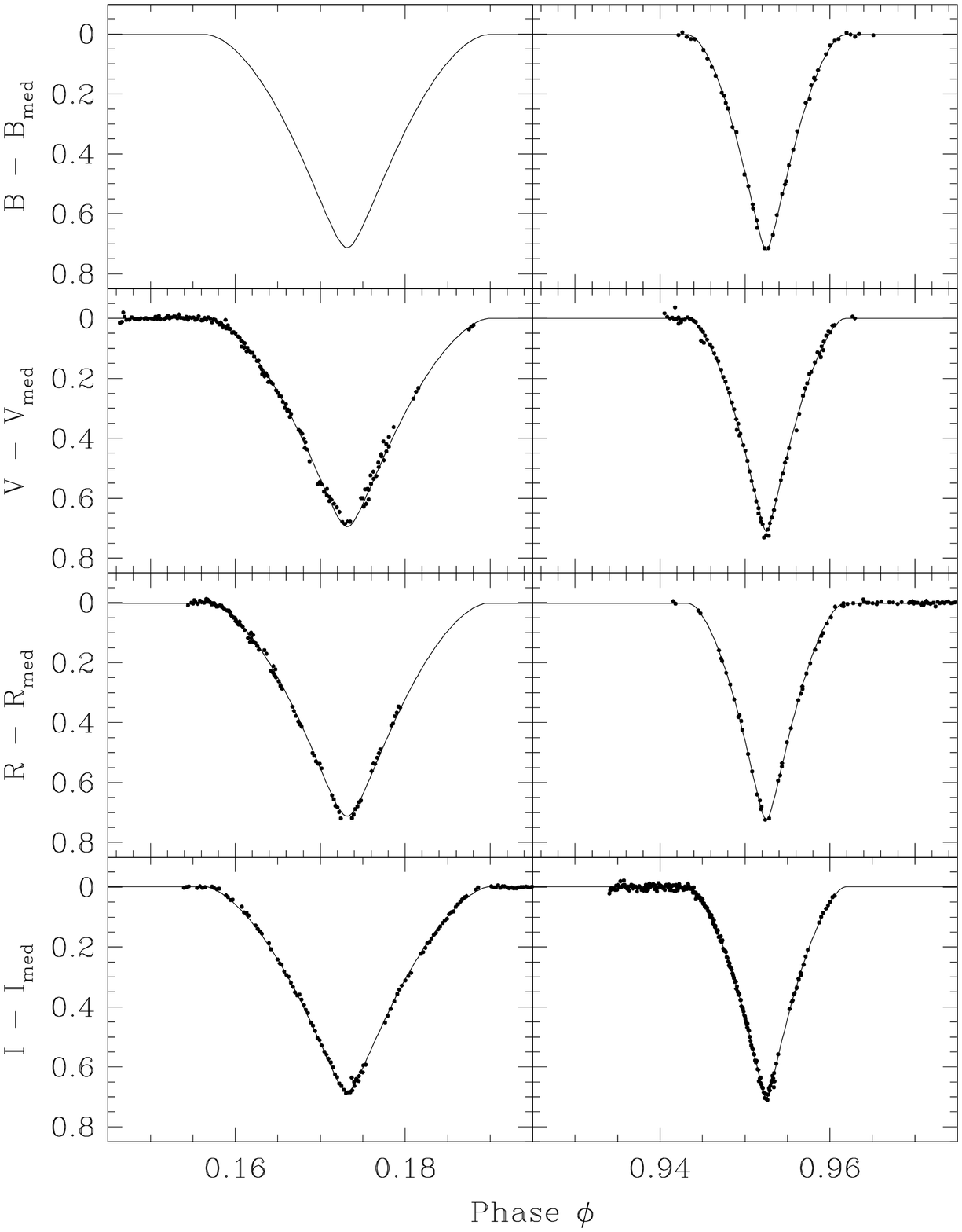}
\caption{$BVRI$ phased light curves for the detached eclipsing binary V2 near
  its eclipses. Model fits are shown.
\label{v2ecl}}
\end{figure}

We trimmed the light curve data down to observations in and near eclipse (see
Fig. \ref{v2ecl}) because of the lack of significant variation at other
phases. As expected for a fairly long period binary, there is no sign of
variation associated with nonsphericity of the stars. We did make zeropoint
adjustments to the photometry (as we did for V375 Cep), but in all cases the
shifts were less than 0.011 mag.

The main result of the analysis is that the two stars have very similar
characteristics. In particular, the mass ratio $q$ is consistent with 1 to
within the $1\sigma$ uncertainty. As a result, we cannot definitively state
which star is the more massive one, and so for the purposes of this paper, we
will define the primary star to be the one eclipsed during the deeper
eclipse. According to the binary star modeling, the primary star is slightly
larger and hotter at about $2\sigma$ and $4\sigma$ levels of significance,
respectively. However, the radius and temperature ratios only differ from 1 by
less than a percent.  These results are supported observationally by the very
long eclipse ingresses and egresses with no sign of totality (in spite of an
inclination found to be be within $0\fdg5$ of $90\degr$), and by the very
similar depths of the eclipses.

\subsection{V1}\label{v1comb}

\begin{figure}
\includegraphics[scale=0.8]{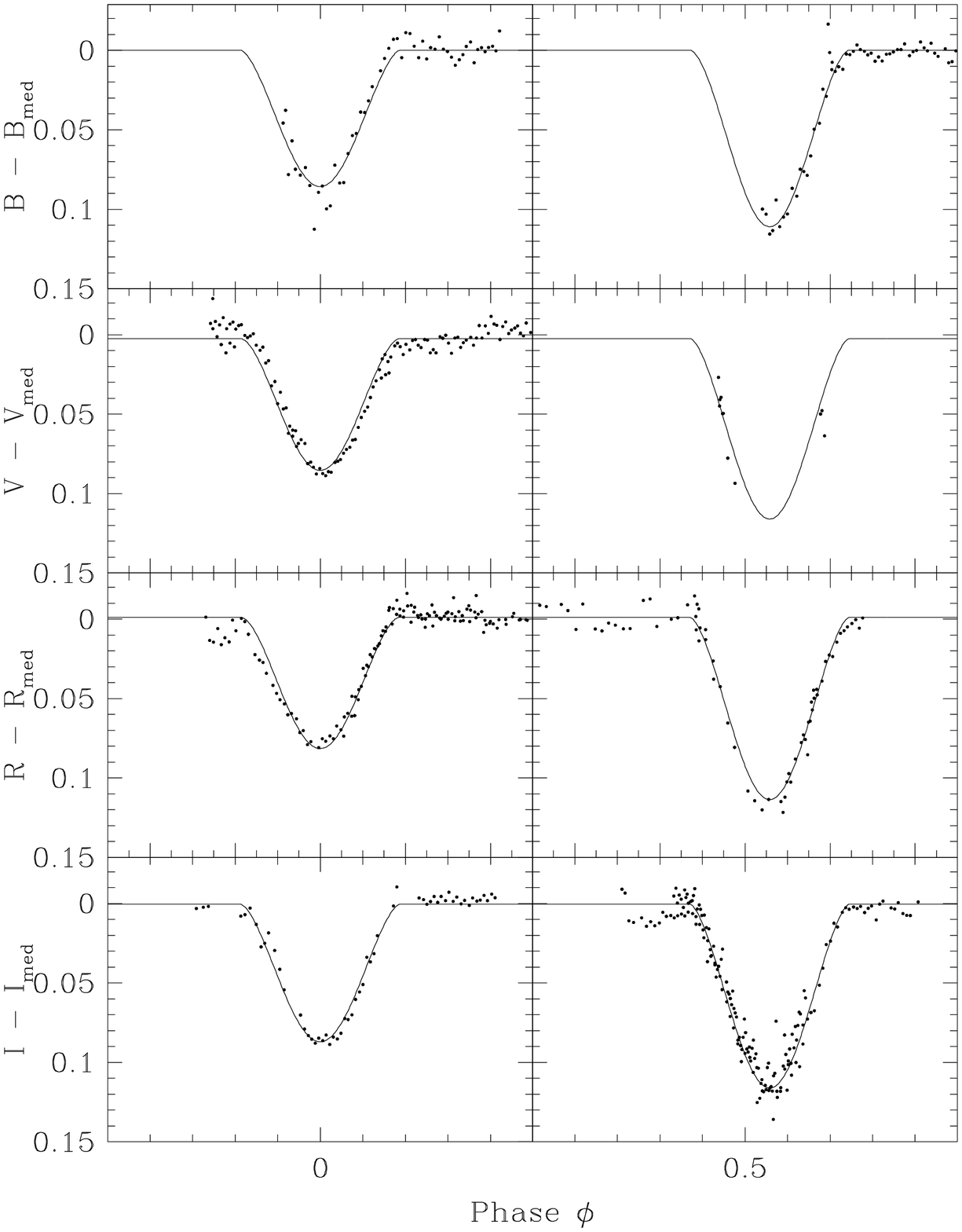}
\caption{$BVRI$ phased light curves for the detached eclipsing binary V1 near
  its eclipses. Model fits are shown.
\label{v1ecl}}
\end{figure}

The combined photometry of this system puts it in the blue straggler portion
of the cluster CMD, and when the components are decomposed, they fall at the
blue end of the distribution of likely cluster stars at the turnoff. However,
our radial velocities clearly identify it as a nonmember. Although we have
only three radial velocity observations, we conducted trial model runs to get
preliminary estimates of the star characteristics. As can be seen in
Fig. \ref{v1ecl}, the light curves show relatively shallow eclipses ($\sim
0.08$ and 0.12 mag), and the secondary eclipse is found at phase $\phi=0.492$,
indicating a slight eccentricity. Binaries with periods shorter than 5 d are
typically found to be circularized even in young populations \citep{mandm}, so
it is worth trying to establish how large the eccentricity is.

In a relatively short period binary like V1, it is not surprising to find some
evidence of spot activity. During most nights of observation, there were few
deviations in the light curve that could be identified with spot
activity. However, on the night of one secondary eclipse (HJD 2455383.8), we
found that the out-of-eclipse level was fainter than was typical in $R$
observations, and there was a difference in the pre- and post-eclipse levels
as well. To correct for this to first order, we applied a zero point shift to
observations from that night to bring the average out-of-eclipse level for
that night into agreement with others.

We then followed a procedure similar to that of binary V2, modeling the radial
velocities and photometry simultaneously. We used the same binary model
parameters with the exception of substituting time of conjunction (primary
eclipse) $t_c$ for time of periastron $t_0$.
$t_c$ is
more directly constrained by observations in our combined dataset for V1.
The model fit indicates that the binary orbit has a very small but significant
eccentricity (the radial velocities are nearly consistent with a circular
orbit), and the long axes of the orbits are almost in the plane of the sky.

\section{Discussion}

\subsection{Mass, Radius, and Age}\label{mr}

\begin{figure}
\plotone{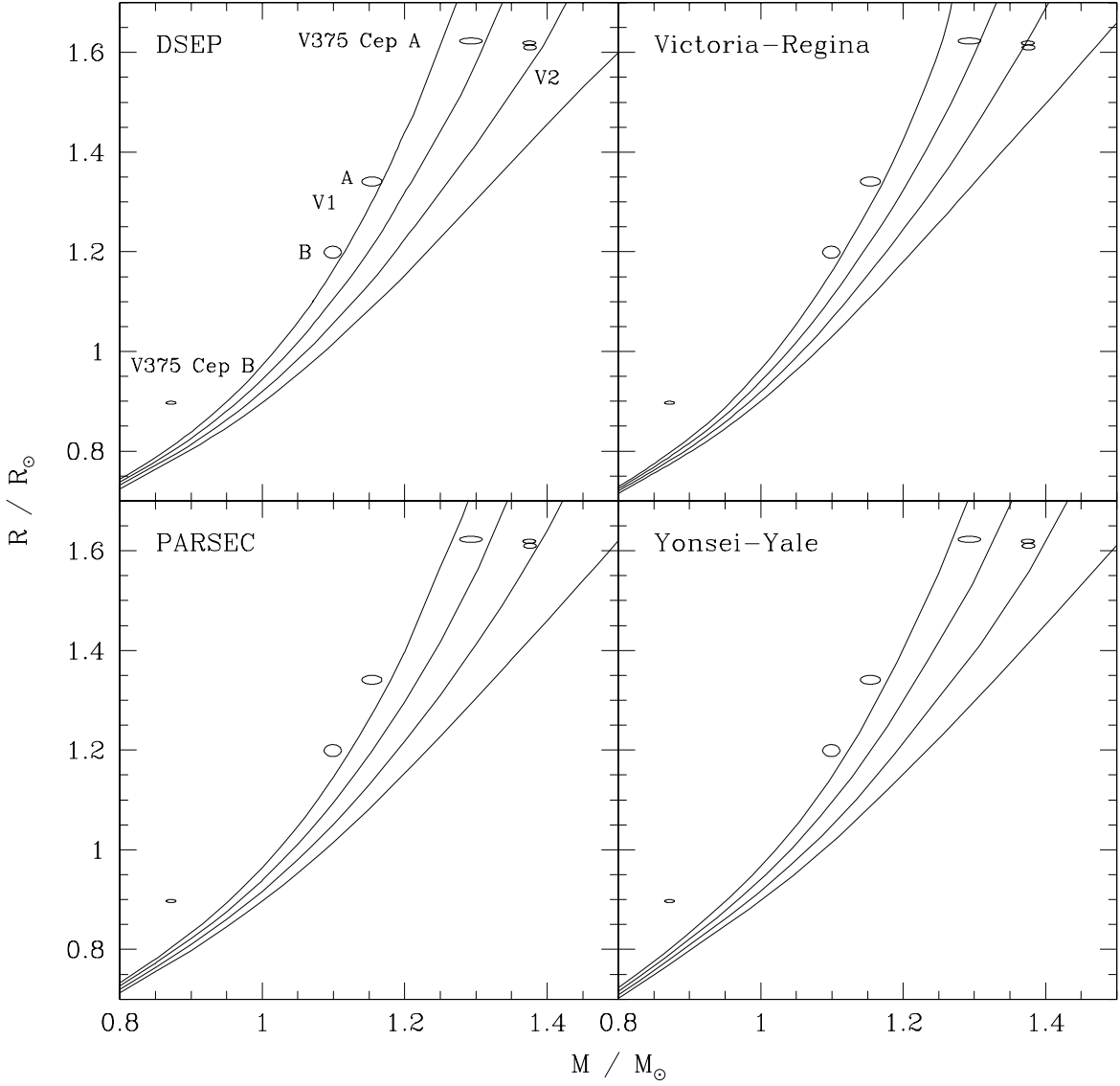}
\caption{$1\sigma$ error ellipses for the eclipsing binaries under
  consideration. In each panel, the isochrones are for ages of 1, 2, 3, and 4
  Gyr (bottom to top). The metal contents are [Fe/H]$=+0.13$ (Victoria-Regina,
  \citealt{vand}), and $+0.14$ (Dartmouth, \citealt{dotter}; PARSEC,
  \citealt{parsec}; and Yonsei-Yale, \citealt{demarque}).
\label{mrplot}}
\end{figure}

The masses and radii for the six stars are plotted in Fig. \ref{mrplot}.
Comparing the results of limb darkening law and model atmosphere runs in Table
\ref{chartab}, there are relatively small ($\sim1$\%) but significant
differences in radius. In the discussion below, we use the results from limb
darkening law runs for their greater ability to fit eclipse ingresses and
egresses. However, it should be remembered that we have not identified the
root cause of the differences.

\subsubsection{V375 Cep}\label{v375disc}

When the components of the two stars in V375 Cep are compared with isochrones
in the $M-R$ plane, we are immediately confronted with several issues.  The
most striking one involves the radius of the lower mass secondary star in the
V375 Cep system. A star of mass $0.87\msun$ should have not have evolved
significantly during the lifetime of a cluster like NGC 7142, but we find that
the star is more than 10\% larger than expected from models.

This kind of behavior has been seen before: \citet{clausen} discuss
well-studied eclipsing binaries in the field containing stars with masses of
$0.80-1.10 \msun$, finding that stars in binaries with short periods
($0.6-2.8$ d) tend to have larger radii and lower temperatures than
predicted. Indicators such as spot-induced photometric variations and X-ray
emission support the idea that stellar activity is related to the radius
discrepancies \citep{torres06}.
Stellar activity is thought to produce magnetic flux tubes that can inhibit
the flow of the convective gas blobs that transport energy to the surface,
forcing the star to grow in size to compensate for the lost transport
capability. Stellar models have been produced that can reproduce such
anomalously large radii via an ad hoc decrease in the mixing length parameter
\citep{chabrier}, or recently via a self-consistent (although
one-dimensional) treatment of the magnetic field \citep{feiden}.

\begin{figure}
\includegraphics[scale=0.7]{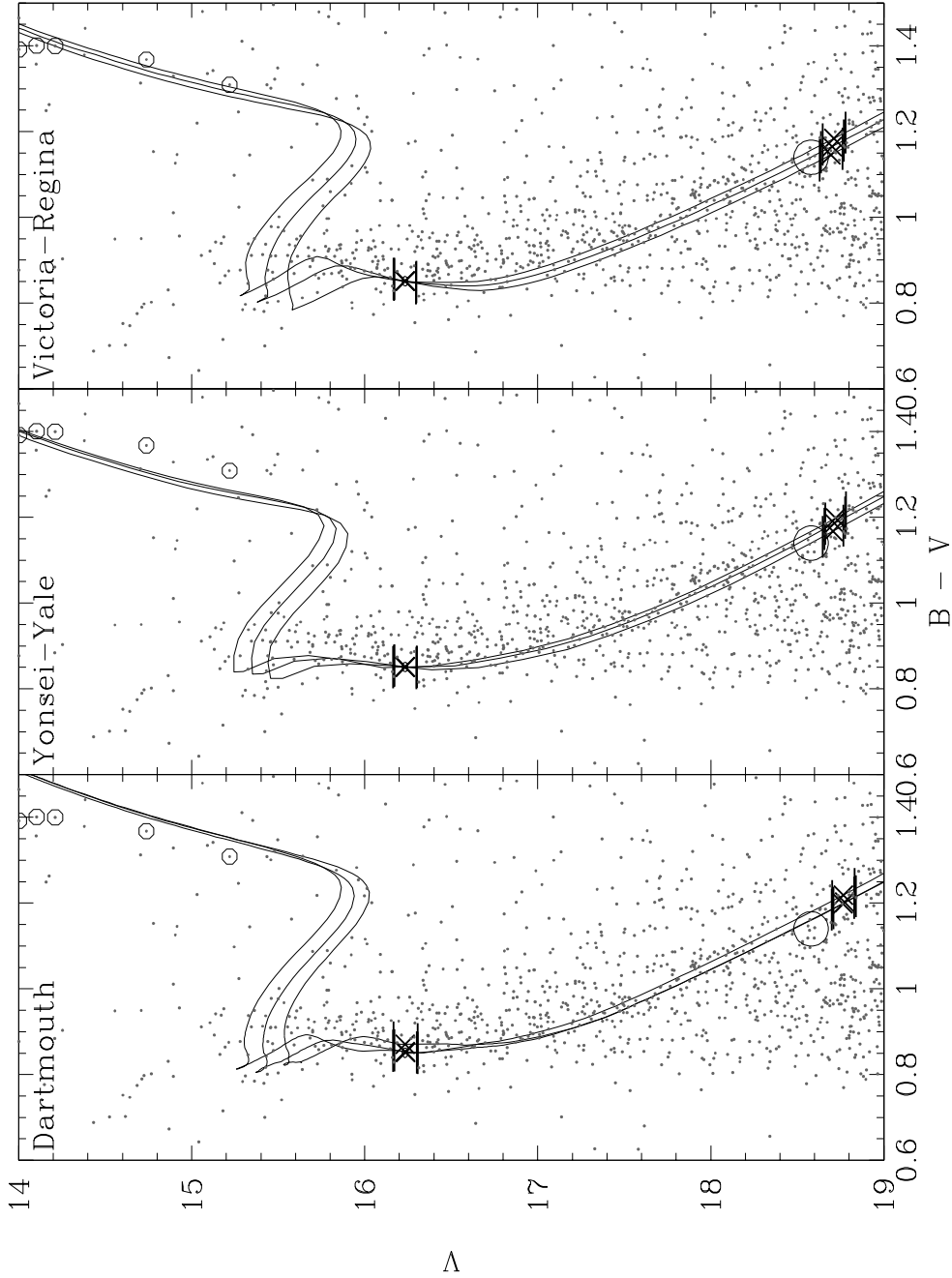}
\caption{($V,B-V$) color-magnitude diagrams for NGC 7142 compared with
  isochrones (ages 3.3, 3.5, and 3.7 Gyr) fitted to the mass and photometry of
  V375 Cep A. The isochrone chemistry is the same as in Fig. \ref{mrplot}. The
  error ellipses for the members of V375 Cep are plotted, along with the
  isochrone points having the same masses (and bars indicating the limits set
  by mass uncertainties). Probable cluster members (identified from
  spectroscopic radial velocities) are shown with small open
  circles.\label{375fit}}
\end{figure}

Because the primary star has a larger mass, its convective
envelope is predicted to be about an order of magnitude smaller in mass than
that of the secondary star. As such, magnetic activity should play a less
important role in influencing the energy transport in the outer layers of the
star, and the radius should be closer to predictions for the cluster age
(although it may still be inflated to a smaller degree). 
The decomposed photometry of the primary star places it toward the blue edge
of the main sequence band for NGC 7142 (see Fig. \ref{375fit}), supporting
the idea that its temperature has not been affected significantly.

FL Lyr \citep{popper86}, V1061 Cyg \citep{torres06}, and EF Aqr \citep{vos}
are three other short-period binaries ($P = 2.1782$ d, 2.3467 d, and 2.8536 d,
respectively) containing inflated secondary stars and primary star masses
($1.218\pm0.016 \msun$, $1.282\pm0.015 \msun$, and $1.244\pm0.008 \msun$,
respectively) similar to that of V375 Cep ($1.288\pm0.017 \msun$). In the
cases of V1061 Cyg and FL Lyr, the primary star radius can be matched with
standard models for reasonable ages of 2.4 and 3.4 Gyr. It should be
understood that this is not conclusive evidence that the primary stars are
free of influences that modify the radius. A modest increase in radius could
be camouflaged as a larger age in both cases. For EF Aqr (which has the
largest orbital period of the three systems), the primary star shows some
signs of being affected. However, the secondary stars can be definitely tagged
as unusual because their radii are significantly larger than could possibly be
expected for reasonable ages at their lower masses.

V375 Cep is a potentially more interesting test case than those field binaries
because the cluster age can be constrained independently using the
color-magnitude diagram, and its orbital period is even shorter.  Taking the
mass and radius of the primary star at face value, an age of about 3.3-3.6 Gyr
is indicated, depending on the model. This is consistent with results from
isochrone fitting in the CMD \citep{sand7142}.  We also measured the
rotational velocities of the two stars from broadening of the spectral
lines. Because the binary has a short period and has circularized, the stars
should have synchronized their spins with the orbit, and the measured
rotational velocities are indeed consistent with synchronous rotation for the
measured stellar radii. In order to check the possibility that magnetic
activity is responsible for the unusually large radius of the secondary, we
looked at several indicators. We see very little evidence of spot-induced
light curve variations unless the variations occur preferentially in the $B$
bandpass. We looked for signs of X-ray emission in archival data from
space-based missions. Although the cluster was observed by {\it XMM-Newton}
(P.I. Verbunt), no source was detected at the position of V375 Cep during a
pointing of more than 10600 s.
So we do not have corroborating evidence that magnetic activity is responsible
for the unusual characteristics of the secondary. A search for emission in the
core of the Ca II H and K lines is probably one of the more promising ways
remaining for proving the presence of such activity.

Unfortunately, the primary star in V375 Cep is so far the only ``normal'' star
that can be used to derive the age of NGC 7142 using the eclipsing binary
technique. Other groups \citep{clausen,torres06} have attempted to derive age
constraints from inflated stars like V375 Cep B using models with reduced
convective mixing length, but this is beyond the scope of this study.  Ideally
a full analysis for this cluster would make use of three or more stars so that
composition questions could be addressed. Helium, for example, is one of the
more substantial unknowns affecting an age analysis, and its abundance for
stars of super-solar metallicity is still somewhat uncertain. However, the
helium abundance can be inferred from the shape of the mass-radius isochrones
if the observational data is sufficiently precise
\citep{brogaard,brogaard2}. Brogaard et al. discussed the old, very metal-rich
cluster NGC 6791, which has a helium abundance significantly above the solar
value. NGC 7142 stars have metal content a little less than halfway between
the Sun and NGC 6791. Until we have additional cluster stars for analysis, we
will implicitly be using helium enrichment laws assumed by the different model
isochrones, meaning that the helium abundance will be super-solar. This
enrichment will have an effect on the age determination if the assumed value
is significantly in error.

\begin{figure}
\includegraphics[scale=0.7]{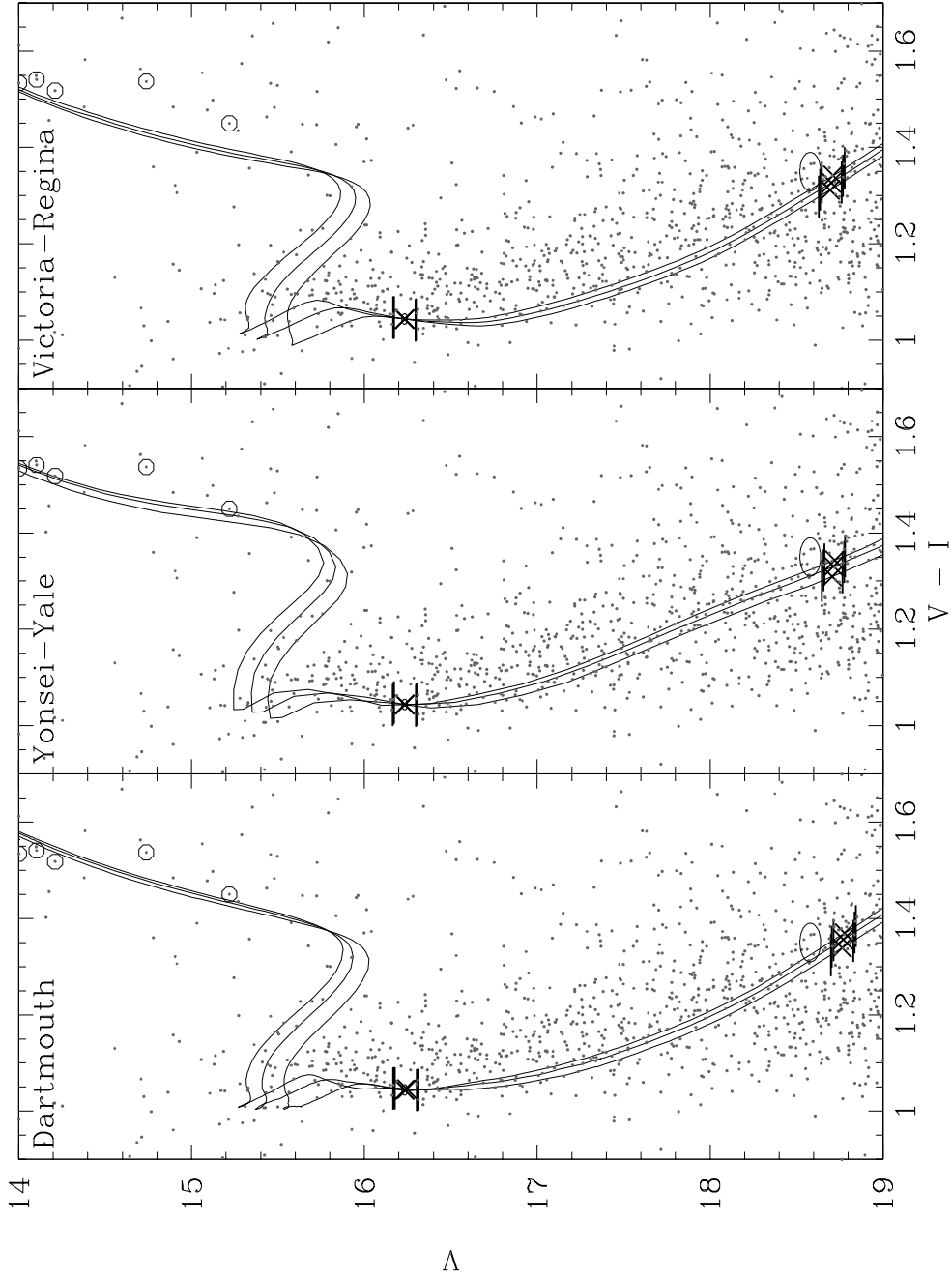}
\caption{($V,V-I_C$) color-magnitude diagrams for NGC 7142 compared with
  isochrones (ages 3.3, 3.5, and 3.7 Gyr) fitted to the mass and photometry of
  V375 Cep A, with symbols as in Fig. \ref{375fit}. \label{375fitvi}}
\end{figure}

Figs. \ref{375fit} and \ref{375fitvi} show a CMD of the cluster with
isochrones pinned to the position of V375 Cep A at the mass measured
here. Generally speaking, isochrones of the age implied by the binary star
analysis are consistent with cluster photometry as well. The details differ
between isochrone sets due to differences in physics. NGC 7142 is in a range
of ages where the physics of the convective core (including core overshooting
and CNO cycle reaction rates) is important.

The photometry of V375 Cep B is consistent with that of a cluster main
sequence star, although the position predicted for a star of its mass from
isochrones is only marginally consistent with the photometry.  Indications
from field binaries \citep{stassun} are that chromospheric activity tends to
increase the stellar radius and decrease effective temperature in a way that
leaves the luminosity unchanged. In short period binaries such as V1061 Cyg
\citep{torres06}, activity induced by forced synchronous rotation also appears
to drive similar changes that keep the luminosity approximately constant.
% Torres finds that Mv agrees for V1061 Cyg, but not R or T (200 k cool)

Because of its position in the CMD, the determination of the mass of V375 Cep
A is essentially a direct measurement of the turnoff mass for the
cluster. Single stars appear to reach slightly bluer colors just before
starting their subgiant branch evolution toward the red giant branch, but V375
Cep A is quite close to bluest point on the main sequence, which is the
traditional definition of the turnoff. Subsequent evolution is comparatively
rapid, and this places a strong upper limit on the age of the
cluster. Isochrones with ages of 4 Gyr or above would require a star of V375
Cep A's mass to have evolved significantly to the blue (and then red). We can
therefore rule out the much greater age ($6.9\pm0.9$ Gyr) determined by
\citet{janes} in their study of the cluster CMD.

As we discussed in a study of eclipsing binaries in the somewhat younger
cluster NGC 6819 \citep{sanda851}, the Dartmouth isochrones \citep{dotter} are
to be preferred among current publicly available isochrones because they
include the most up-to-date inputs for physics that affects the evolution of
turnoff-mass stars. Since that paper, the PARSEC models \citep{parsec} have
been revised, and contain similar input physics. One important difference
between the Dartmouth and PARSEC models and most others is the inclusion of an
improved nuclear reaction rate for the CNO cycle reaction
$^{14}$N$(p,\gamma)^{15}$O. In addition, stellar model calculations typically
include a varying amount of convective core overshooting for stars with masses
around that of V375 Cep A. The amount (expressed in units of the pressure
scale height $H_P$) is typically ramped up from zero at a lower mass limit
($1.1 \msun$ for the Dartmouth models) to a maximum value at a high mass limit
($0.2 H_P$ at $1.3 \msun$). Both physics effects have minimal effects on the
color-magnitude diagram except near the cluster turnoff and subgiant branch,
where the details of central hydrogen exhaustion in the stars significantly
influence the shape of the isochrones. With the possibility of nailing down
the isochrones at the position of one or more binary stars, we therefore have
leverage to test the physics of the stellar cores. This test would be stronger
in NGC 7142 if the effects of differential reddening and field star
contamination could be reduced. Until then, the indications are that different
isochrone sets can reproduce the cluster turnoff in a qualitative sense.

Before leaving the discussion of this binary star, we use the stars to
calculate a distance modulus for the cluster. From the measured radius and
effective temperature, we can calculate the bolometric luminosity.  For the
temperature, we have used the spectroscopic estimate in order to avoid
uncertainties associated with the cluster reddening. After applying a
theoretical bolometric correction \citep{vandc}, we derive $M_V$ and the
distance modulus. The primary star provides the best estimate
[$(m-M)_V=12.86\pm0.07$] because its photometry and its effective temperature
are better constrained, but the measurement from the secondary star is
completely consistent [$(m-M)_V=12.86\pm0.15$] if the star's temperature is
derived from the effective temperature of the primary and the temperature
ratio from the light curve fits using the analytic limb darkening law. These
measurements are in nice agreement with our previous determination using
the color-magnitude diagram ($12.96\pm0.24$; \citealt{sand7142}), but are of
higher precision and effectively independent of any need for reddening
estimates.

\subsubsection{V2}\label{v2mr}

In the case of the V2 system, we find that the masses and radii (as well as
their temperatures) agree to within 1\%, and the masses are consistent with
being equal to within the $1\sigma$ uncertainties. Both stars appear to have
evolved significantly and equally in radius, and these facts imply that the
characteristics of these stars were set early on in their evolution, and have
remained unchanged. Nearly equal mass binaries are commonly found in the field
and in cluster environments (see \citealt{qdist} and references therein) in
agreement with hydrodynamical simulations of fragmentation
\citep[e.g.][]{bate} during the star formation process.  It is difficult to
imagine a process (such as stable mass transfer) that could have forced the
masses of the stars to become equal {\it after} birth without circularizing the
orbits. Even if one could, the differences in the prior rates of evolution
for the stars (in other words, how much of the central hydrogen had been
processed to helium) would produce differences in radius. The equality of the
stars along with their eccentric orbits imply that they have evolved
undisturbed since their formation.

If V2 was a current or former member of NGC 7142 and had the same chemical
composition, we should expect the stars to fall on the same isochrone as the
primary star in V375 Cep. The characteristics of the stars of V2 differ from
the isochrone that passes through V375 Cep A at about the $4.5\sigma$ level,
appearing to be about 1.5 Gyr younger. This is the most convincing evidence
that the stars in V2 are not cluster members --- the mass and radius pairs
imply age and/or chemical composition that is significantly different than the
cluster. A calculation of the distance modulus for this binary star
using the photometric temperature estimate returns $(m-M)_V=12.55\pm0.09$,
which is significantly smaller than found for the cluster member V375 Cep.

\subsubsection{V1}

The binary star V1 has essentially zero probability of cluster membership
based on its system velocity and the signs that its reddening is lower than
the other binaries. Because the stars are part of a relatively short period
binary, we checked to see whether there were signs of radius inflation, as
there is for V375 Cep B (see the earlier discussion). Although the metallicity
of the binary has not yet been determined, the two stars have positions in the
$M-R$ diagram that are consistent with being on the same isochrone. Both stars
are likely to still have convective envelopes but with smaller mass than the
Sun's, and the period of the binary is larger than found for other systems
with seemingly inflated stars. V636 Cen \citep{clausen} is an interesting
comparison, having a slightly shorter period (4.28 d) than V1, but having a
secondary star of lower mass ($0.87 \msun$) with a more massive convective
envelope.  Based on a simple interpretation of the stellar activity
hypothesis, the two stars should be expected to show small or no radius
inflation.

The small eccentricity ($e=0.038$) that is detected is also of some interest.
The circularization timescale for the binary is around a Gyr according
to the formulation in \citet{zahn} for stars with convective envelopes ---
less than, but of similar magnitude to, the age of the binary. It is therefore
plausible that the circularization process has not been completed for this
binary. If this binary does not have a third orbiting object that is
maintaining the eccentricity, we could be seeing the final stages of
circularization as brought on by the evolutionary expansion of the stars. Their
expansion, even over the last Gyr, has significantly decreased the
circularization timescale by about a factor of 2.

Because the metallicity of the binary is not known, it is not possible to
derive a precise age, but the indication is that the binary is slightly older
($\sim 1$ Gyr) than NGC 7142. To put this differently, 
any isochrone that connects the two stars in V1 does not pass through the error
ellipse for the primary star in V375 Cep. This provides more
evidence that the system is not a member of NGC 7142, if any was needed.

\section{Conclusions}

We have studied three detached eclipsing binary stars that were discovered
near the turnoff of the open cluster NGC 7142 in the color-magnitude
diagram. From multiple lines of evidence, we conclude that the V375 Cep system
is the only one of the three that is a cluster member. The measured mass and
radius of the primary star of V375 Cep support an age of 3.3-3.6 Gyr for the
cluster. We are also able to compute a distance modulus for the cluster
[$(m-M)_V=12.86\pm0.07$] that is mostly independent of estimates of the
cluster reddening.

V375 Cep is a short period binary, however, and this appears to be responsible
for the abnormally large radius of the secondary star.  Because the binary has
total eclipses of the secondary star, we can accurately disentangle the
photometry of the two stars.  Because the binary is a member of the cluster,
we can use the photometry for other cluster stars to judge whether the
components of V375 Cep have experienced a color/temperature shift. The primary
star is found toward the blue end of the cluster main sequence band, so its
surface does not appear to have been affected by the interactions with its
companion. The secondary star shows clear evidence that its radius has been
inflated, and there is some marginal evidence that it is slightly redder in
the $(V-I_c)$ color than predicted by models. Higher precision observations of
the secondary eclipse will be needed to prove this point more definitively.

In order to further test the connection between magnetic activity and the
inflated radius of the secondary, additional targeted observations are called
for. Our spectroscopy has provided rotational velocities for the stars that
are consistent with synchronous rotation, supporting the possibility of
rotationally-induced activity, but there are not yet strong tests of magnetic
activity in the system. X-ray emission might reveal activity in the system,
although the archived {\it XMM-Newton} integration did not reveal V375 to be a
significant X-ray source.  Because V375 Cep is more distant than
commonly-studied field binaries, deeper observations would be challenging.
A search for emission in the cores of the Ca II H and K lines \citep{clausen}
would seem to be the best next test.

\acknowledgments This work has been funded through grant AST 09-08536 from the
National Science Foundation to E.L.S. We would like to thank the Director of
Mount Laguna Observatory (P. Etzel) for generous allocations of observing
time. Infrastructure support for the observatory was generously provided by
the National Science Foundation through the Program for Research and Education
using Small Telescopes (PREST) under grant AST 05-19686. 

The Hobby-Eberly Telescope (HET) is a joint project of the University of Texas
at Austin, the Pennsylvania State University, Stanford University,
Ludwig-Maximilians-Universitat Munchen, and Georg-August-Universitat
Gottingen.  The HET is named in honor of its principal benefactors, William
P. Hobby and Robert E. Eberly.  This research made use of the SIMBAD database,
operated at CDS, Strasbourg, France; and the NASA/IPAC Infrared Science
Archive, which is operated by the Jet Propulsion Laboratory, California
Institute of Technology, under contract with the National Aeronautics and
Space Administration.

\begin{figure}
\includegraphics[scale=0.8]{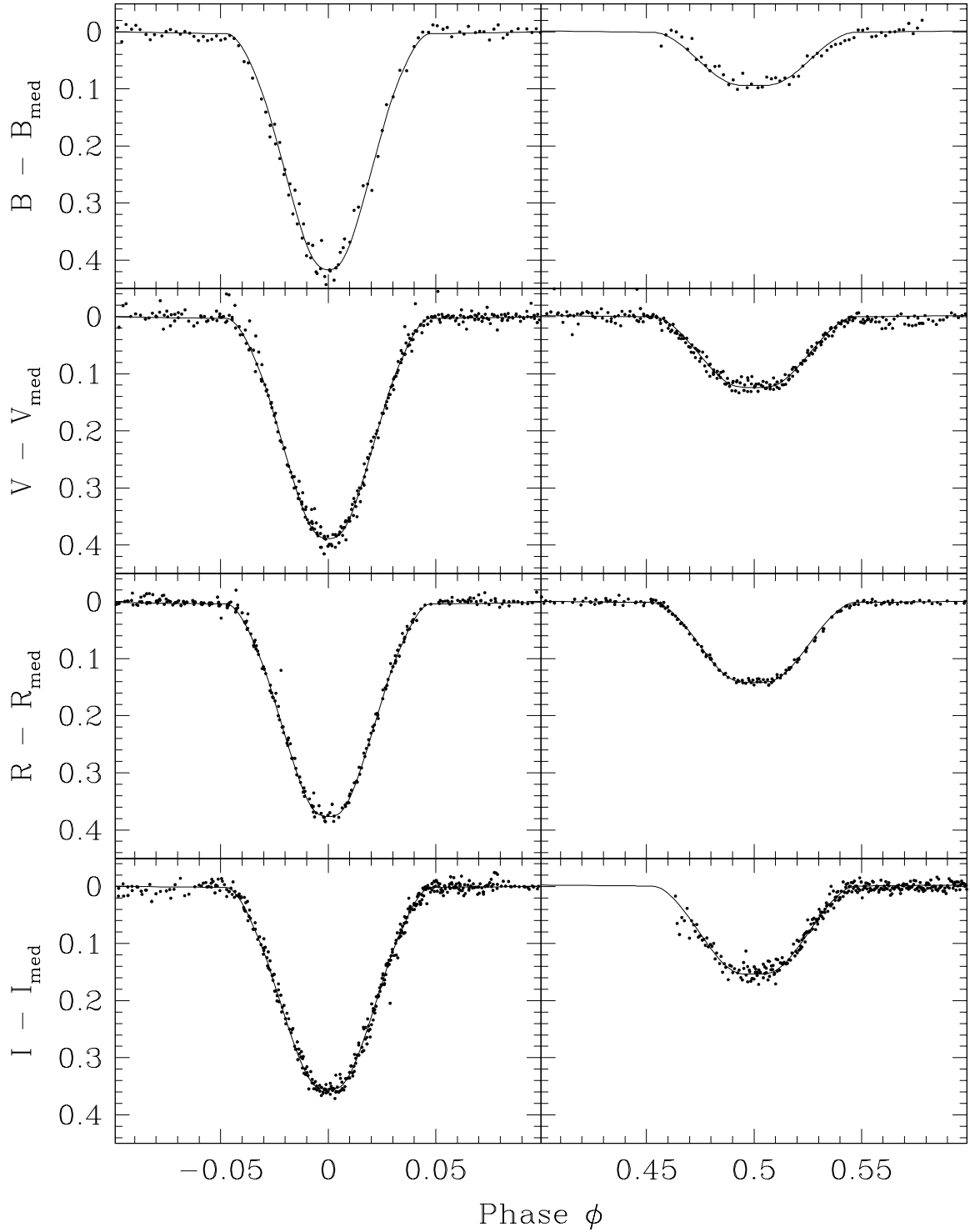}
\caption{$BVRI$ phased light curves for the detached eclipsing binary V375 Cep
  near its eclipses. Model fits (employing analytic limb darkening laws) are
  shown.
\label{v375ecl}}
\end{figure}

\begin{deluxetable}{clclc}
\tablewidth{0pt}
\tablecaption{Additional Photometry at Mount Laguna Observatory}
\tablehead{\colhead{Date} &
\colhead{Filters} & \colhead{mJD Start\tablenotemark{a}} & \colhead{$N$}}
\startdata
2011 9 Aug. & $BR$ & 5783.656 & 38,22 \\
2011 28 Aug. & $VR$ & 5802.629 & 25,71 \\
2011 14 Oct. & $VR$ & 5849.816 & 27,3 \\
2011 30 Nov. & $R$ & 5896.566 & 21 \\
\enddata
\label{datestab}
\tablenotetext{a}{mJD = HJD - 2450000.}  
\end{deluxetable}

\begin{deluxetable}{crrrrr}
\tablewidth{0pt}
\tablecaption{Radial Velocity Measurements}
\tablehead{\colhead{UT Date} &
\colhead{mJD} & \colhead{$v_A$ (\kms)} & \colhead{$\sigma_{A}$} & \colhead{$v_B$ (\kms)} & \colhead{$\sigma_B$}}
\startdata
\multicolumn{6}{c}{V1:}\\
20100625 &  5372.91397 &    55.9 &   0.7 &   $-94.2$&  1.0\\
20100728 &  5405.81053 &    66.7 &   0.8 &  $-102.6$&  1.3\\
20100731 &  5408.82025 &  $-83.3$&   1.0 &     52.0 &  1.0\\
\multicolumn{6}{c}{V2:}\\%05/12 revision
20091015 &  5119.59672 &  $-70.0$&   2.2 &   $-13.6$&  2.5\\
20091022 &  5126.60325 &   $-3.2$&   1.2 &   $-83.3$&  1.8\\
20091125 &  5160.55958 &     4.1 &   1.8 &   $-90.9$&  2.2\\
20100916 &  5455.65953 &    29.1 &   1.5 &  $-114.5$&  2.2\\
20100916 &  5455.77462 &    37.8 &   1.9 &  $-122.3$&  2.3\\
20101007 &  5476.64403 &  $-46.4$&   1.4 &   $-33.3$&  4.1\\
20101009 &  5478.65485 &  $-66.0$&   0.9 &   $-20.0$&  1.3\\
20101010 &  5479.64021 &  $-71.8$&   0.9 &   $-14.0$&  1.1\\
20101011 &  5480.61719 &  $-76.1$&   1.3 &    $-8.9$&  1.5\\
20101019 &  5488.61755 &    27.2 &   1.1 &  $-112.7$&  1.4\\
20101103 &  5503.58119 &    53.1 &   1.1 &  $-139.0$&  1.8\\
20101105\tablenotemark{b} &  5505.57139 &         &       &   $-72.9$&  0.9\\
20110707 &  5749.89305 &  $-80.2$&   0.9 &    $-4.6$&  1.2\\
20110710 &  5752.86970 &    17.1 &   0.8 &  $-101.8$&  1.1\\
20110809 &  5782.78284 &  $-61.0$&   1.1 &   $-24.0$&  1.4\\
20110823 &  5796.75684 &  $-80.3$&   1.8 &    $-4.8$&  2.1\\
20110827 &  5800.73788 &    56.8 &   1.5 &  $-143.3$&  1.9\\
20110908 &  5812.73498 &  $-79.7$&   1.1 &    $-7.1$&  1.5\\
20110925 &  5829.65263 &  $-63.7$&   1.5 &   $-22.6$&  2.1\\
20110927 &  5831.63652 &    52.4 &   2.1 &  $-138.4$&  2.7\\
\multicolumn{6}{c}{V375 Cep:}\\%revised 06/12
20080905 &  4714.73837 &    18.1 &   2.6 &  $-158.4$&  3.0 \\
20080926 &  4735.67830 &    27.6 &   1.0 &  $-163.4$&  2.6 \\
20080929 &  4738.68081 & $-101.4$&   1.1 &     24.3 &  1.8 \\
20080930 &  4739.66833 &   $-5.3$&   0.8 &  $-112.5$&  2.1 \\
20081017 &  4756.63643 &    33.4 &   5.0 &  $-178.8$&  3.3 \\
20081018 &  4757.57776 & $-135.5$&   0.7 &     76.1 &  1.4 \\
20081031 &  4770.58112 & $-103.3$&   0.8 &     34.4 &  5.0 \\
20081104 &  4774.56837 & $-134.3$&   1.3 &     77.8 &  1.9 \\
20081106 &  4776.58653 & $-139.0$&   1.0 &     79.7 &  2.0 \\
20081107 &  4777.56955 &    39.5 &   1.7 &  $-179.1$&  2.6 \\
20081107 &  4777.60328 &    34.4 &   2.0 &  $-179.7$&  3.3 \\
20081118 &  4788.56133 &  $-38.8$&   1.2 &   $-73.3$&  8.0 \\
20081120 &  4790.55131 &  $-16.2$&   0.9 &   $-97.3$&  2.6 \\
20081122 &  4792.55191 &     6.2 &   1.2 &  $-132.8$&  3.3 \\
20100909 &  5448.68348 & $-133.8$&   2.6 &     74.4 &  4.2 \\
20100913 &  5452.71345 & $-134.5$&   0.9 &     78.2 &  3.0 \\
20100915 &  5454.72577 & $-121.4$&   2.3 &     54.4 &  4.5 \\
20101003 &  5472.67656 &    41.4 &   1.8 &  $-180.0$&  3.4 \\
20101005 &  5474.65060 &    38.3 &   1.8 &  $-178.5$&  3.8 \\
20101024 &  5493.58271 &    37.5 &   2.4 &  $-177.4$&  3.9 \\
20101105 &  5505.58498 &  $-33.2$&   3.6 &   $-61.3$& 10.0 \\
20110926 &  5830.68360 & $-138.2$&   0.9 &     80.6 &  2.3 \\  
\enddata
\label{spectab}
\tablenotetext{a}{mJD = HJD - 2450000.}  
\tablenotetext{b}{Observation during eclipse.}
\end{deluxetable}

\begin{deluxetable}{lccccccc}
\tablewidth{0pt}
\tabletypesize{\scriptsize}
\tablecaption{Photometry of V1, V2, and V375 Cep}
\tablehead{\colhead{Star} & \colhead{$B$} & \colhead{$V$} & \colhead{$R_C$} & \colhead{$I_C$} & \colhead{$J$} & \colhead{$H$} & \colhead{$K_s$}}
\startdata
V1 & \\
Combined & $15.695\pm0.009$ & $14.864\pm0.010$ & & $13.825\pm0.010$ & $13.096\pm0.027$ & $12.757\pm0.035$ & $12.671\pm0.029$ \\% 2MASS ID 21443286+6545263
Primary & 16.29 & 15.46 & & 14.43 & 13.70 & 13.37 & 13.28 \\
Secondary & 16.63 & 15.79 & & 14.75 & 14.02 & 13.67 & 13.59 \\
\hline
V2 & \\
Combined & $16.107\pm0.008$ & $15.310\pm0.009$ & & $14.300\pm0.009$ & $13.661\pm0.032$ & $13.342\pm0.033$ & $13.227\pm0.037$ \\
\hline
V375 Cep & \\
Combined & $16.992\pm0.009$ & $16.115\pm0.010$ & & $15.035\pm0.011$ & $14.326\pm0.031$ & $13.877\pm0.040$ & $13.814\pm0.048$ \\
Ecl. Depth & $0.092\pm0.008$ & $0.118\pm0.004$ & $0.140\pm0.001$ & $0.154\pm0.007$ & & & \\% from medians on nights of eclipse
Primary & $17.084\pm0.012$ & $16.233\pm0.011$ & & $15.189\pm0.013$ & & & \\
Secondary & $19.72\pm0.09$ & $18.58\pm0.04$ & & $17.23\pm0.05$ & & & \\
\enddata
\label{phottab}
\end{deluxetable}

\begin{deluxetable}{ccc}
\tablewidth{0pt}
\tablecaption{Photometric Minima}
\tablehead{\colhead{Eclipse} & \colhead{Filter} & \colhead{mJD\tablenotemark{a}}}
\startdata
\multicolumn{3}{c}{V1}\\
\hline
P & $R$ & $53639.9356\pm0.0006$\\
P & $I$ & $54657.7878\pm0.0005$\\
S & $V$ & $54678.7786\pm0.0008$\\
S & $B$ & $55355.7864\pm0.0008$\\
S & $R$ & $55383.7955\pm0.0007$\\
P & $I$ & $55390.8320\pm0.0009$\\
S & $I$ & $55411.8168\pm0.0014$\\
\hline
\multicolumn{3}{c}{V2} \\
\hline
P & $R$ & $53639.6889\pm0.0002$\\
P & $I$ & $54656.9770\pm0.0011$\\
P & $V$ & $55423.8558\pm0.0011$\\
P & $V$ & $55517.7581\pm0.0002$\\
P & $I$ & $55736.8673\pm0.0003$\\
S & $I$ & $55755.9707\pm0.0004$\\
P & $B$ & $55783.8178\pm0.0003$\\
S & $V$ & $55802.9238\pm0.0026$\\
S & $V$ & $55849.8720\pm0.0011$\\
\hline
\multicolumn{3}{c}{V375 Cep}\\
\hline
P & $V$ & $46650.484^{+0.005}_{0.020}$\\%
P & $V$ & $47442.81\pm0.025$\\%
P & $BV$ & $47469.659\pm0.004$\\% fit using a rough chi^2 algorithm
S & $R$ & $53594.9749\pm0.0066$\\%gap,few obs after min
S & $R$ & $53596.8816\pm0.0003$\\
S & $R$ & $53598.7923\pm0.0005$\\
P & $R$ & $53599.7467\pm0.0002$\\
S & $R$ & $53600.7011\pm0.0006$\\
P & $R$ & $53637.9396\pm0.0003$\\
P & $R$ & $53639.8496\pm0.0002$\\
S & $R$ & $53640.8054\pm0.0004$\\
P & $R$ & $53641.7594\pm0.0002$\\
P & $B$ & $54630.9733\pm0.0024$\\
S & $B$ & $54633.8450\pm0.0016$\\
P & $B$ & $54634.7949\pm0.0007$\\
P & $I$ & $54655.8013\pm0.0003$\\
S & $I$ & $54656.7542\pm0.0008$\\
P & $I$ & $54657.7114\pm0.0002$\\%a litle asym,few obs before min
S & $V$ & $54675.8491\pm0.0012$\\
P & $V$ & $54676.8071\pm0.0005$\\
S & $V$ & $54677.7630\pm0.0007$\\
P & $V$ & $54678.7170\pm0.0004$\\
P & $V$ & $55121.7629\pm0.0005$\\
P & $V$ & $55146.5934\pm0.0009$\\%few obs before minimum
S & $V$ & $55372.8853\pm0.0012$\\%noticable asymmetry
\enddata
\label{mintab}
\tablenotetext{a}{mJD = HJD - 2400000.}  
\end{deluxetable}

\begin{deluxetable}{l|cc|cc|c}
\rotate
\tablewidth{0pt}
\tablecaption{Characteristics of the Eclipsing Binaries}
\tablehead{\colhead{Parameter} & \multicolumn{2}{c}{V375 Cep} & \multicolumn{2}{c}{V2} & \colhead{V1} \\
 & \colhead{Limb Darkening} & \colhead{Atmospheres} & \colhead{Limb Darkening} & \colhead{Atmospheres} & }
\startdata
$\gamma$ (km s$^{-1}$) & \multicolumn{2}{c}{$-49.86 \pm 0.05$} & $-42.64\pm0.02$ & $-42.66$ & $-17.16^{+0.03}_{-0.09}$ \\%7/12 run
$q$ & \multicolumn{2}{c}{$0.676\pm0.004$} & $1.0001\pm0.0033$ & 0.9992 & $0.951\pm0.004$ \\%7/12 run
$K_A$ (km s$^{-1}$) & \multicolumn{2}{c}{$89.13\pm0.24$} & $69.92\pm0.20$ & 69.91 & $80.81\pm0.30$ \\%7/12 run
$v_{rot}(A)$ (km s$^{-1}$) & \multicolumn{2}{c}{$42^{+2}_{-4}$ (set)} & & & \\% checked with Matt 8/5/12
$v_{rot}(B)$ (km s$^{-1}$) & \multicolumn{2}{c}{$20^{+10}_{-5}$ (set)} & & & \\
$t_0-2450000$ & 3599.74650 & 3599.74651 & 5502.85043 & 5502.8507 & \\
$\sigma(t_0)$ & $\pm0.00007$ & $\pm0.00008$ & $\pm0.00008$ & &  \\
$t_c-2450000$ & & & $$ & & $3639.9062\pm0.0003$ \\
$P$ (d) & 1.90968257 & 1.90968252 & 15.6505950 & 15.6505954 & 4.6690576 \\
$\sigma(P)$ (d) & $\pm0.00000016$ & $\pm0.00000018$ & $\pm0.0000016$ & & $\pm0.0000007$ \\
$i$ ($\degr$) & $85.34^{+0.02}_{-0.05}$ & $85.39^{+0.03}_{-0.02}$ & $89.703\pm0.008$ & 89.650 & $83.38\pm0.02$ \\
$e$ & \multicolumn{2}{c}{0 (set)} & $0.52165\pm0.00002$ & 0.52158 & $0.0379^{+0.0031}_{-0.0004}$\\
$\omega$ ($\degr$) & \multicolumn{2}{c}{90 (set)} & $326.692\pm0.003$ & 326.696 & $285.0^{+0.2}_{-1.2}$\\
$R_A / a$ & $0.1937\pm0.0003$ & $0.1958\pm0.0004$ & $0.04381\pm0.00011$ & 0.04350 & $0.0877\pm0.0009$\\
$R_A / R_B$ & $1.8098\pm0.0018$ & $1.840\pm0.002$ & $1.0022\pm0.0030$ & 0.9847 & $1.140^{+0.010}_{-0.021}$\\
$R_B / a$ & $0.10704\pm0.00020$ & $0.10644^{+0.00014}_{-0.00022}$ & $0.04372\pm0.00012$ & 0.04417 & $0.0769\pm0.0007$\\
$(R_A + R_B) / a$ & $0.3008\pm0.0005$ & $0.3023\pm0.0006$ & $0.08753\pm0.00012$ & 0.08767 & $0.1647\pm0.0003$\\
$T_B / T_A$ & $0.8268\pm0.0008$ & $0.8122\pm0.0012$ & $0.9969\pm0.0005$ & 0.9944 & $0.983\pm0.002$\\
$M_A$ ($\msun$) & $1.288\pm0.017$ & $1.288\pm0.017$ & $1.377\pm0.009$ & 1.379 & $1.147\pm0.012$ \\%7/23 run
$M_B$ ($\msun$) & $0.871\pm0.008$ & $0.871\pm0.008$ & $1.377\pm0.009$ & 1.378 & $1.090\pm0.013$\\%7/23 run
$R_A$ ($\rsun$) & $1.623\pm0.006$ & $1.642\pm0.007$ & $1.616\pm0.005$ & 1.605 & $1.338\pm0.008$\\%7/23 run
$R_B$ ($\rsun$) & $0.897\pm0.003$ & $0.893\pm0.004$ & $1.613\pm0.005$ & 1.630 & $1.190\pm0.008$\\%7/23 run
$\log g_A$ (cgs) & $4.129\pm0.003$ & $4.120\pm0.003$ & $4.160\pm0.003$ & 4.166 & $4.244\pm0.007$ \\%7/23 run
$\log g_B$ (cgs) & $4.473\pm0.002$ & $4.478\pm0.002$ & $4.162\pm0.003$ & 4.152 & $4.324\pm0.007$ %7/23 run
\enddata
\label{chartab}
\end{deluxetable}

\end{document}